\documentclass[a4paper,twocolumn,amsmath,amssymb,pra,footinbib,floatfix,superscriptaddress]{revtex4}
\usepackage[utf8]{inputenc}

\usepackage{graphicx}
\usepackage{dcolumn}
\usepackage{bm}
\usepackage{multirow}
\usepackage{mathdots}
\usepackage{amssymb}
\usepackage{amsmath}
\usepackage{hyperref}
\usepackage{color}
\usepackage{soul}
\usepackage{physics}
\usepackage{MnSymbol}
\usepackage{float}
\usepackage{mathtools}
\usepackage{slashed}

\definecolor{red}{rgb}{0.7,0,0}
\definecolor{green}{rgb}{0.,0.35,0.}
\definecolor{blue}{rgb}{0.2,0.2,0.7} 
\definecolor{black}{rgb}{0.15,0.15,.15}
\def\mel#1#2#3{\langle #1 | #2 | #3 \rangle}

\DeclareMathOperator{\sign}{sign}	

\def\({\left(}
\def\){\right)}
\def\<{\langle}
\def\>{\rangle}
\def\{{\lbrace}
\def\}{\rbrace}
\def\({\left(}
\def\){\right)}
\def\[{\left[}
\def\]{\right]}

\newcommand{\sx}{\sigma^x}

\newcommand{\sz}{\sigma^z}

\newcommand{\lp}{\left(}
\newcommand{\rp}{\right)}

\usepackage{tikz}
\usetikzlibrary{shapes.geometric,arrows}
\makeindex

\begin{document}

\title{Entanglement in non-critical inhomogeneous quantum chains}

\author{Nadir Samos Sáenz de Buruaga}
\affiliation{Instituto de Física Teórica UAM/CSIC, Universidad
  Autónoma de Madrid, Cantoblanco, Madrid, Spain}

\author{Silvia N. Santalla}
\affiliation{Dto. Física \&\ GISC, Universidad Carlos III de
  Madrid, Spain}

\author{Javier Rodríguez-Laguna}
\affiliation{Dto. Física Fundamental, Universidad Nacional de
  Educación a Distancia (UNED), Madrid, Spain}

\author{Germán Sierra}
\affiliation{Instituto de Física Teórica UAM/CSIC, Universidad
  Autónoma de Madrid, Cantoblanco, Madrid, Spain}

\begin{abstract}
We study an inhomogeneous critical Ising chain in a transverse field
whose couplings decay exponentially from the center.
In the strong inhomogeneity limit we apply Fisher's renormalization group
to show that the ground state is formed by concentric singlets
similar to those of the rainbow state of the XX model.
In the weak inhomogeneity limit we map the model to a
massless Majorana fermion living in a hyperbolic spacetime,
where, using  CFT techniques, we derive the entanglement entropy that is violated linearly. 
We also study an inhomogeneous non-critical Ising model that
for weak inhomogeneity is mapped to a massive Majorana
fermion, while for  strong inhomogeneity regime it exhibits
trivial and non-trivial topological phases and a separation
between regions with high and low entanglement. We also present the entanglement Hamiltonian
of the model. 
\end{abstract}

\date{July 25, 2021}

\maketitle


\section{Introduction} 
\label{sec:intro}

The study of entanglement in quantum many-body systems
\cite{Amico.08,Calabrese.09b,Laflorencie.16,Roy.19} has proven to be
an excellent way to advance in the understanding of quantum matter
\cite{Wen.book}. Given a pure state $\ket{\Psi}$ of a system and a
bipartition into two subsystems, $A\cup B$, all the information
concerning quantum correlations between these parts is contained in
the reduced density matrix $\rho_A=\Tr_B\ket{\Psi}\bra{\Psi}$. The
most important measure of entanglement is the von Neumann entropy,
$S_A=- \Tr\rho_A\log\rho_A$, that vanishes if and only if the
subsystems are disentangled. The low energy states of local quantum
Hamiltonians are expected to satisfy the so-called area law, which
asserts that the entanglement entropy (EE) of a block is bounded by
the size of its boundary
\cite{Sredniki.93,Wolf.08,Hastings.06,Eisert.10}. This property holds
for one dimensional gapped Hamiltonians under certain assumptions
\cite{Hastings.06}, but is violated in critical Hamiltonians described
by a conformal field theory (CFT), where the EE grows logarithmically
with the subsystem size and is proportional to the central charge
\cite{Holzhey.94,Vidal.03,Calabrese.04,Calabrese.09}.

The study of entanglement in spatially inhomogeneous systems has
attracted recently considerable interest. For example, the EE of local
Hamiltonians with random couplings exhibits a logarithmic behaviour,
presenting some similarity with the CFT result
\cite{Refael.04,Refael.04b,Laflorencie.05,Fagotti.11,Ramirez.14,Ruggiero.16}.
Other interesting cases include the engineered trapping potentials for
ultracold atoms which reduce the boundary effects
\cite{Nishino.09,Campostrini.10,Dubail.17,Murciano.19}, the interplay
between quantum gravity and strange metals within the SYK model
\cite{Sachdev.93,Rosenhaus.19}, or the quantum simulation of the Dirac
vacuum on a curved spacetime using optical lattices
\cite{Boada.11,Laguna.17,Mula.21}. In some cases, it is possible to
obtain the GS of these systems using renormalization group (RG)
schemes such as the Dasgupta-Ma procedure
\cite{Dasgupta.80}. Conversely, the RG can help us design certain
lattice models, such as the {\em rainbow state} (RS), which is the GS
of an XX spin-chain whose couplings decay exponentially from the
center towards the edges, and which presents a linearly growing EE
between its two halves
\cite{Vitagliano.10,Ramirez.14b,Alkurtass.14,Ramirez.15,Samos.19,Laguna.16,
  Laguna.17b,Tonni.18,MacCormack.18,Samos.20,AA18,Alba.19}. The system
is described in terms of an inhomogeneity parameter $h$, associated to
the exponential decay. In the strong inhomogeneity regime, the GS is a
nested set of Bell pairs, which can also be described as a concentric
singlet state \cite{Vitagliano.10,Alkurtass.14}. In the weak
inhomogeneity regime the rainbow chain corresponds to a free Dirac
theory on a (1+1)D anti-de-Sitter space-time
\cite{Laguna.17b,MacCormack.18}. Thus, the EE can be obtained as a
deformation of well known results from CFT, showing that the
logarithmic growth maps into a linear one. Moreover, there is a smooth
crossover between the weak and strong inhomogeneity regimes.

In addition to the EE, bipartite entanglement can be characterized by
other magnitudes. The density matrix $\rho_A$ can be written as

\begin{equation}
  \rho_A\propto e^{-\mathcal{H}_A},
\end{equation}
where $\mathcal{H}_A$ is called the entanglement Hamiltonian (EH)
associated to a block $A$ within a quantum state
\cite{Li_Haldane.08,BW.75,BW.76,Peschel.09,Cardy.16,Eisler.19,Eisler.20}. Even
if the original state is translationally invariant, the EH usually
represents an inhomogeneous system, and it has been characterized for
several interesting systems, employing e.g. the corner matrix
formalism (CTM) \cite{Peschel.09,Eisler.20} or CFT results
\cite{Cardy.16}. Moreover, the EH of certain inhomogeneous systems has
been described using the RS as a benchmark, showing that it can be
understood as a {\em thermofield double}: each half of the system can
be approximated by a homogeneous system at a finite temperature that
depends on the inhomogeneity level $h$ \cite{Tonni.18}. The spectrum
of the EH is called the entanglement spectrum (ES), which can provide
interesting information e.g. regarding the existence of
symmetry-protected topological phases (SPT)
\cite{Kitaev.01,Pollman10,Fidkowski2011,Turner2011,W11a,P12}.

The aim of the present work is twofold. First, to characterize the
emergence of a rainbow state as a ground state of an inhomogeneous
transverse field Ising (ITF) Hamiltonian, when the couplings and the
external fields are allowed to decay in a certain way, by mapping it
to a (1+1)D massless Majorana field on a curved space-time. Then, we
will describe the structure of the model away from the critical point,
showing that it reduces to a massive Majorana field in the same
setup. Moreover, we shall also consider the relation between our model
and the Kitaev chain \cite{Kitaev.01}.

This article is organized as follows. In section \ref{sec:critical} we
introduce an inhomogeneous version of the ITF model and describe its
entanglement structure. The strong inhomogeneity regime is discussed
by means of RG schemes, and the weak inhomogeneity regime is
characterized via field theory methods. In Sec. \ref{sec:noncrit} we
propose a variation of the previous model by adding a new parameter
which shifts it away from the critical point, and we describe its
entanglement properties in the strong and weak inhomogeneity
regimes. Finally, we present a brief discussion of our conclusions and
prospects in Sec. \ref{sec:conclusions}.


\section{The Ising Rainbow State}
\label{sec:critical}
Let us consider an inhomogeneous ITF open spin $1/2$ chain with an even number of sites $N=2L$ whose Hamiltonian is defined as:

\begin{equation}
\label{eq:ham_ising}
H_I=-\sum_{m=-L+1/2}^{L-3/2} J_m\sz_m\sz_{m+1}-\sum_{m=-L+1/2}^{L-1/2}\Gamma_m\sx_m,
\end{equation}
where $\sx$ and $\sz$ are Pauli matrices. Notice that the spins are
indexed by half-odd integers for later convenience, $m =-L+1/2,\dots,
L-1/2$. We shall apply a Jordan-Wigner transformation and write
Eq. \eqref{eq:ham_ising} in terms of Dirac fermions $c^\dagger_m$
which satisfy the usual anti-commutation relations
$\{c^\dagger_m,c_{n}\}=\delta_{mn}$, and then decompose them in terms
of Majorana fermions

\begin{equation}
   c_m=\frac{1}{2}(\alpha_m+i\beta_m), 
   \label{eq:cfermions}
\end{equation}
that satisfy the anti-commutation relations
$\{\alpha_m,\alpha_{n}\}=\{\beta_m,\beta_n\}=2\delta_{mn}$ and
$\{\alpha_m,\beta_n\}=0$. In terms of these Majorana fermions
Eq. \eqref{eq:ham_ising} reads:

\begin{align}
\label{eq:major_ising}
  H=&-i\Bigl(\sum_{m=1/2}^{L-1/2}\Gamma_m\(\alpha_m\beta_m+\alpha_{-m}\beta_{-m}\)\\
  &+ \sum_{m=1/2}^{L-3/2}J_m\(\beta_m\alpha_{m+1}+\beta_{-(m+1)}\alpha_{-m}\)+J_{-1/2}\beta_{-1/2}\alpha_{1/2}\Bigr)\nonumber,
\end{align}

\begin{figure}[H]
\includegraphics[width=80mm]{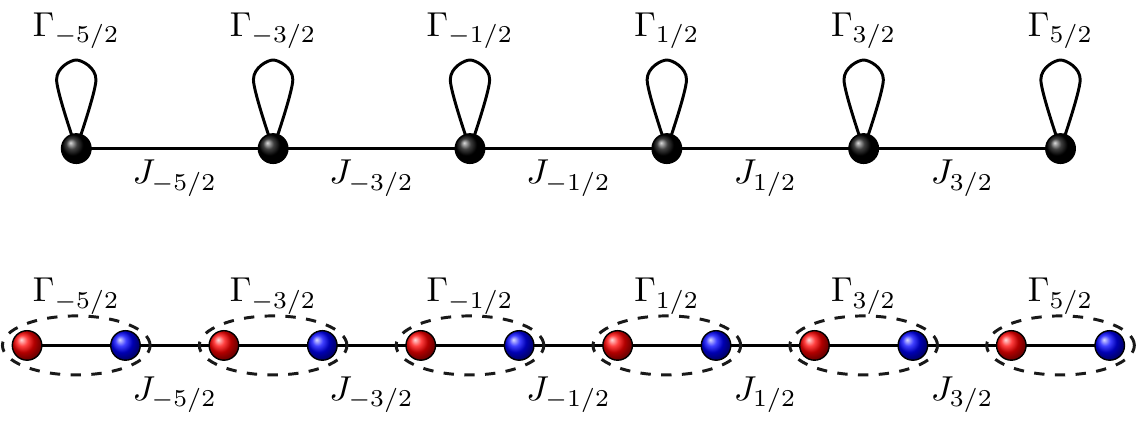}\\
\caption{Spin (top) and Majorana fermion (bottom) representations of
  the inhomogeneous ITF model. The red (blue) points correspond to
  $\alpha$ ($\beta$) Majorana fermions.}
\label{fig:chain}
\end{figure}

Notice that the same system is described by $2L$ spins and $4N$
Majorana fermions. In Fig. \ref{fig:chain} we present an schematic
representation of the model in terms of spins (top) and Majorana
fermions (bottom). The transverse field $\Gamma_m$ couples two
Majorana fermions with the same index $(\alpha_m,\beta_m)$, while the
coupling constants $J_m$ link Majorana fermions with different indices
$(\beta_m,\alpha_{m+1})$. The dashed lines that encircle the Majorana
fermions represent the Dirac fermions $c_m$ in
Eq. \eqref{eq:cfermions}.

Notice that if $J_m=\Gamma_m=1$ for all $m$ we recover the critical
ITF model whose low-energy behaviour is described by the two
dimensional Ising CFT with central charge $c=1/2$. In addition, if
$J_m=0$ for all $m$, the ground state becomes a trivial product state
built upon the physical fermions $c_m$. On the contrary, if
$\Gamma_m=0$ for all $m$, the Majorana fermions placed at the edges of
the Majorana chain, $\alpha_{-L+1/2}$ and $\beta_{L-1/2}$, do not
appear in the Hamiltonian, Eq. \eqref{eq:major_ising}. Moreover, they
correspond to Majorana zero modes and the GS belongs to the
topologically non-trivial phase of the Kitaev model
\cite{Kitaev.01}.

Let us consider both the spin and the Majorana fermion Hamiltonians,
Eqs. \eqref{eq:ham_ising} and \eqref{eq:major_ising}, under the
following choice of coupling constants $J_m$ and $\Gamma_m$:

\begin{eqnarray}
J_m &=& \begin{cases}
  e^{-2h|m+1/2|} & \text{if } m\neq -1/2, \nonumber\\
  e^{-h/2} & m=-1/2,\\
\end{cases}\\
\Gamma_m&=&e^{-2h|m|},
\label{eq:hop_ising}
\end{eqnarray}
where $h\geq0$ is the inhomogeneity parameter. Notice that for $h>0$
the intensity of the couplings decreases from the center towards the
edges, with $J_{-1/2}$ corresponding to the strongest coupling. Also,
the system is symmetric under reflections around the central bond,
satisfying $J_{-(m+1)}=J_m$ and $\Gamma_m=\Gamma_{-m}$. In the
remainder of this section, we shall describe the strong ($h\gg 1$) and
weak ($h\ll 1$) inhomogeneity regimes.

\subsection{Strong Inhomogeneity} 
\label{sub:strong_inhomogeneity}

In the limit $h\gg1$ we can characterize the GS of
\eqref{eq:major_ising} using the strong disorder renormalization
scheme (SDRG) developed by Fisher for the ITF
\cite{Fisher.94,Fisher.95}. It proceeds by finding the strongest
interaction coupling, either $\Gamma$ or $J$, which gets sequentially
decimated. If it corresponds to a magnetic field, $\Gamma_i$, the
$i$-th spin is integrated out, leaving the system with one spin less
and a new coupling term between the spins $i-1$ and $i+1$,

\begin{equation}
  \tilde{J}_{i-1}\sigma^z_{i-1}\sigma^z_{i+1},\qquad
  \text{ with } \; \;  \tilde{J}_{i-1}=\frac{J_{i-1}J_{i}}{\Gamma_{i}}.
\label{eq:rg_J}
\end{equation}
On the other hand, if the coupling $J_i$ is the strongest interaction
at a given RG step, the spins $i$ and $i+1$ get renormalized into a
single spin with effective Hamiltonian

\begin{equation}
  \tilde{\Gamma}_i\sigma^x_i,\qquad
  \text{ with } \tilde{\Gamma}_i=\frac{\Gamma_i\Gamma_{i+1}}{J_i}.
\label{eq:rg_G}
\end{equation}
Notice that renormalizing a $J$ term entangles two neighboring spins,
while the renormalization of a $\Gamma$ term freezes that spin along the
direction of the magnetic field, and decouples it from the chain.

It can be shown that the fusion rules of Majorana fermions correspond
to the $SU(2)_2$ quantum group
\cite{Bonesteel.07,Pachos.17,Pachos.book,Sierra.book,Nayak.08}. The
non-univocal fusion rule

\begin{equation}
  \frac{1}{2}\cross\frac{1}{2}=0+1,
  \label{eq:nabelian_frule}
\end{equation}
corresponds to the pairing of two Majorana fermions, for instance
$\alpha_m\beta_m$, which in terms of of Dirac fermions is $2\lp
c^\dagger_m c_m-1/2\rp$, see Eq. \eqref{eq:cfermions}, and we can
attach the fusion channel $0$ ($1$) to the $-1$ ($+1$)
eigenvalue. Indeed, notice that Eq. \eqref{eq:nabelian_frule} reminds
the composition of two $1/2$ spins. Thus, we may call the less
energetic channel a {\em generalized singlet state}
\cite{Bonesteel.07,Fidkowsky.08}, and the other channel as a {\em
  generalized triplet} (albeit there is no $S_z$ degeneracy). Notice
that while the $1/2$ spins obey the $SU(2)$ algebra and the singlet
states span the total $S_z=0$ Hilbert space sector, Majorana fermions
obey the $SU(2)_{k=2}$ algebra and the generalized singlet state spans
the Hilbert space sector for the fusion channel $0$.

With this parallelism in mind, we can devise an SDRG specially suited
for an inhomogeneous Majorana chain \cite{Motrunich.01,Devakul.17}, as
it is done in Appendix \ref{sec:dasgupta_majos}. At each RG step, the
two Majorana fermions linked through the strongest coupling (notice
that in terms of Majorana fermions the $J$ and $\Gamma$ terms are
equivalent) are fused into their less energetic channel, forming a
generalized singlet state or {\em bond}, and leaving a renormalized
coupling between their closest neighbors. This scheme is completely
equivalent to Fisher's RG, Eqs. \eqref{eq:rg_J} and
\eqref{eq:rg_G}. In this case, the SDRG becomes analogous to the
Dasgupta-Ma technique for spin-1/2 XX chains \cite{Dasgupta.80}, for
which it can be proved that the bonds never cross \cite{Laguna.16}.

\bigskip
 
\begin{figure}
\includegraphics[width=80mm]{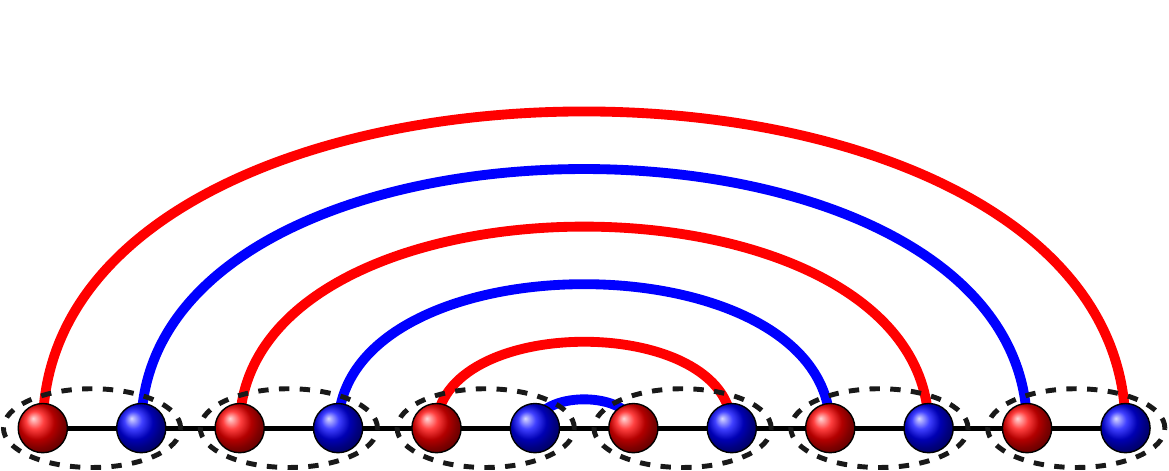}\\
\includegraphics[width=80mm]{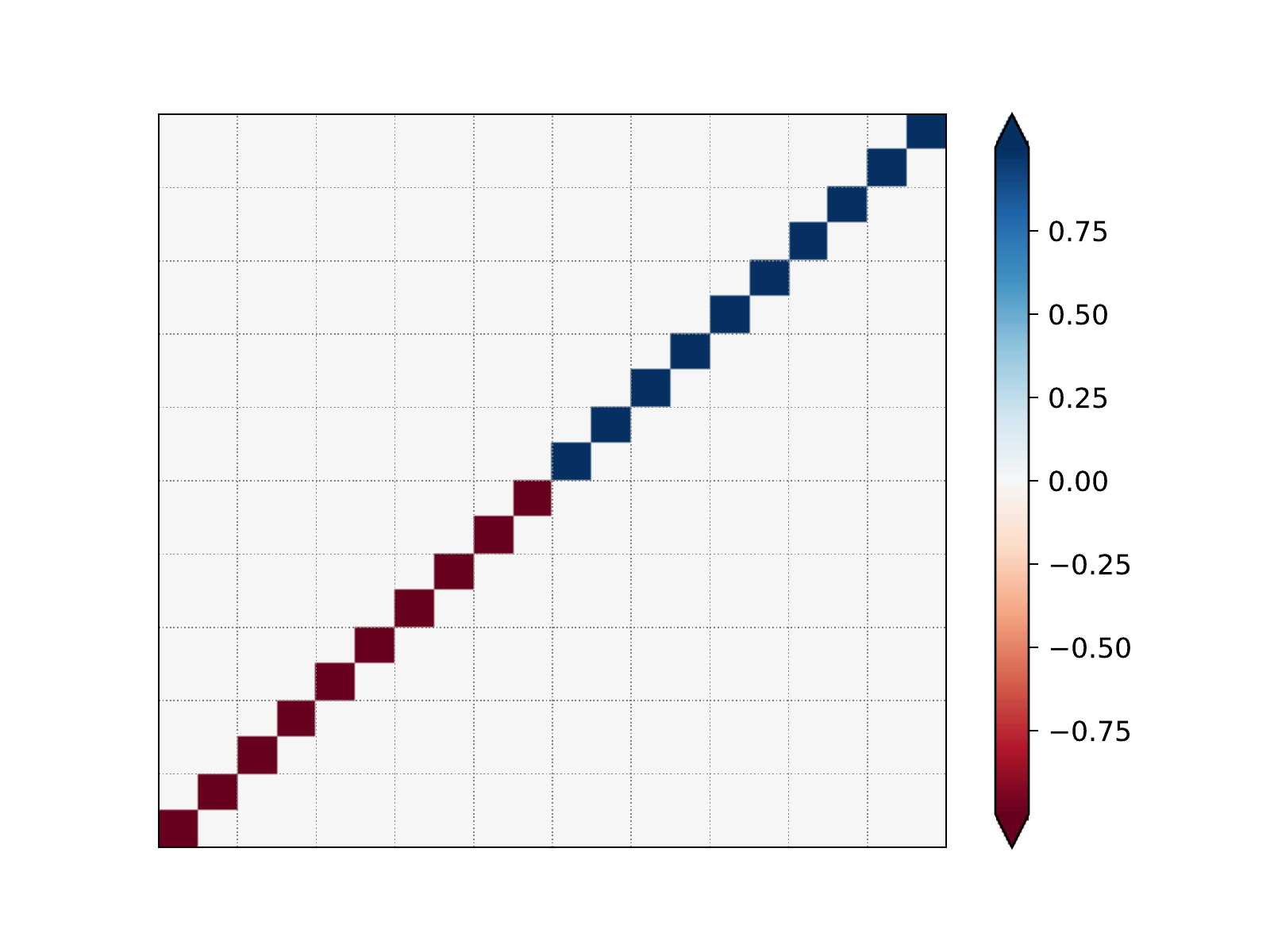}\\
\caption{Top: schematic representation of the outcome of the RG for a
  chain of 6 spins. Blue bonds stand for $b$-type Dirac fermions,
  while red ones represent $d$-type fermions, see
  Eq. \eqref{eq:bdfermions}, which can also be considered as
  generalized singlet states corresponding to the less energetic
  fusion channel. Bottom: covariance
  matrix of the RS for $N=20$ and $h=10$. Notice that the non-zero
  elements are in the anti-diagonal, as it is dictated by the
  structure of the $b$ and $d$ fermions.}
\label{fig:majos}
\end{figure}

Let us apply this RG scheme to the Majorana Hamiltonian given in
Eq. \eqref{eq:major_ising}. The first Majorana pair to be decimated is
$(\beta_{-1/2},\alpha_{1/2})$, because $J_{-1/2}$ is the strongest
coupling. Hence, these two Majorana fermions fuse into a Dirac fermion,

\begin{equation}
b_{1/2}=\frac{1}{2}\lp\beta_{-1/2}+i\alpha_{1/2}\rp,
\label{eq:b12}
\end{equation}
which becomes decoupled. Using Eq. \eqref{eq:rg_G} we can find an
effective Hamiltonian with $2(N-1)$ Majorana fermions, whose new
central term $\tilde{\Gamma}_{1/2}\alpha_{-1/2}\beta_{1/2}$ is given
by

\begin{equation}
\tilde{\Gamma}_{1/2}=\frac{\Gamma_{-1/2}\Gamma_{1/2}}{J_{-1/2}}=e^{-\frac{3h}{2}}.
\end{equation} 
The strongest coupling is now $\tilde{\Gamma}_{1/2}$. We apply the RG
again, and the decimated Majorana fermions fuse into a Dirac fermion,

\begin{equation}
d_{1/2}=\frac{1}{2}\lp\alpha_{-1/2}+i\beta_{1/2}\rp.
\label{eq:d12}
\end{equation}
The new effective Hamiltonian of $2(N-2)$ Majorana fermions has a
central term $\tilde{J}_{3/2}$ which is given by Eq. \eqref{eq:rg_J},

\begin{equation}
\tilde{J}_{3/2}=\frac{J_{-3/2}J_{1/2}}{\tilde{\Gamma}_{1/2}}=e^{-\frac{5h}{2}},
\end{equation}
which is again the strongest coupling in the chain. Given the symmetry
of the coupling constants, Eq. \eqref{eq:hop_ising}, all RG steps
decimate the central pair of Majorana fermions, fusing them
alternatively into $b$ and $d$ Dirac fermions. Hence, the ground
state, that we shall call the Majorana rainbow state
$\ket{\text{MRS}}$, is annihilated by the following Dirac operators:

\begin{equation}
b_m\ket{\text{MRS}}=0,\quad d_m\ket{\text{MRS}}=0,\quad
m=\frac{1}{2},\dots,L-\frac{1}{2}, \\
\label{eq:rstate}
\end{equation}
with

\begin{equation}
b_m=\frac{1}{2}\lp\beta_{-m}+i\alpha_m\rp,\quad
d_m=\frac{1}{2}\lp\alpha_{-m}+i\beta_m\rp.
\label{eq:bdfermions}
\end{equation}
$\ket{\text{MRS}}$ is a concentric generalized singlet state, shown in
Fig. \ref{fig:majos}.

It is worth to compare the Majorana rainbow state
Eq. \eqref{eq:rstate} with its Dirac counterpart, which emerges as the
GS of the inhomogeneous XX chain and its fermionic version
\cite{Ramirez.14b,Ramirez.15,Laguna.16,Laguna.17b}. This state can be
seen as a singlet state of concentric bonding and antibonding
operators,

\begin{equation}
\ket{\text{RS}_{\text{XX}}}=\prod_{m=1/2}^{L-1/2}(b^-_m)^\dagger
(b^+_m)^\dagger\ket{0},
\label{eq:rsxx}
\end{equation}
with

\begin{equation}
\begin{cases}b^+_m=\frac{1}{\sqrt{2}}\lp c_m+c_{-m}\rp\\
b^-_m=\frac{1}{\sqrt{2}}\lp c_m-c_{-m}\rp.\end{cases}
\end{equation}
The alternation between bonding and antibonding is due to the
non-local nature of the Jordan-Wigner transformation, since each
long-distance coupling acquires a phase related to the number of
fermions contained in it. In spin language, the XX model
$\ket{RS_{\text{XX}}}$ is formed by spin-$1/2$ concentric singlets. The
alternation between bonding and antibonding in Eq. \eqref{eq:rsxx} is
therefore similar to that of $b$ and $d$ Dirac fermions in
Eq. \eqref{eq:bdfermions}.

\bigskip

Let us compute the EE of a subsystem $A$, with length $L_A$, for the
Majorana RS, Eq. \eqref{eq:rstate}. The EE of any partition of a
ground state formed by $SU(2)$ singlet states can be estimated by
counting the number of bonds which cross the partition boundary, and
multiplying by $\log(2)$. The procedure is the same when we deal with
generalized singlet states. The EE of any subsystem $A$ can be
estimated by counting the number of bonds which cross the partition
boundary and multiplying by $\log d$ \cite{Bonesteel.07}, where
$d=\sqrt{2}$ is the associated quantum dimension to the algebra
$SU(2)_2$.

Alternatively, the EE of a Gaussian state can be obtained from its
covariance matrix (CM), $\mathcal{C}$,

\begin{equation}
  \mathcal{C}_{ab}=\expectationvalue{[\gamma_a,\gamma_b]},
\end{equation}
where we have arranged the Majorana operators in a vector form
$\bm{\gamma}^T=(\alpha_{-L+1/2},\beta_{-L+1/2},\dots,\alpha_{L-1/2})$. In
Appendix \ref{sec:covariance} we provide a brief derivation of this
expression. The structure of the GS obtained through the decimation
procedure shows up in the CM, as we can see in Fig. \ref{fig:majos}
(b). The EE of a subsystem $A$ with size $L_A$ can be computed through
the eigenvalues $\pm\lambda_k$, $k\in\{1,\cdots,L_A/2\}$ of the
appropriate restriction $\mathcal{C}_A$ of the CM \cite{Peschel.03},
through

\begin{equation}
  S_A=-\sum_{k=1}^{L_A} \nu_k\log\nu_k, \qquad
  \nu_k=\frac{1}{2}\lp1+\lambda_k\rp.
    \label{eq:ee_cov}
\end{equation}  
We can now compute the EE of a lateral block of the system,
$A_\ell=\{-L+1/2\dots-L+1/2+2\ell\}$, with $\ell=1,\cdots,L$. Notice
that a block with an odd number of Majorana operators has no
physical sense. Thus, $A_\ell$ must contain an even number of Majorana
fermions, which correspond to the physical fermions (dotted boxes) or
the spins (black balls) of Fig. \ref{fig:chain}. We can obtain the EE
by counting the number of bonds $n_b$ (reds and blues) that $A_\ell$
cuts in Fig. \ref{fig:majos} and multiply it by $\log d$.

\begin{equation}
S_A=n_b\log\sqrt{2}.
\label{eq:ent_recipe}
\end{equation}
Hence, we have that the EE of the MRS grows linearly,

\begin{equation}
S(A_\ell)=2\ell\log{\sqrt{2}}=\ell\log2,
\label{eq:ent_rb}
\end{equation}
and the maximal EE corresponds to the half chain block
$S(A_L)=L\log2$.


\subsection{Weak Inhomogeneity Regime} 
\label{sub:weak_inhomogeneity_regime}

In this section we shall consider the GS of Eq. \eqref{eq:ham_ising}
with couplings given by Eq. \eqref{eq:hop_ising}, in the low
inhomogeneity regime, $h\ll1$. The equations of motion associated to
the lattice Hamiltonian in the Heisenberg picture are given by
$i\partial_t\alpha_{\pm m}=[H,\alpha_{\pm m}]$ and
$i\partial_t\beta_{\pm m}=[H,\beta_{\pm m}]$. Using
Eq. \eqref{eq:major_ising} we have

\begin{eqnarray}
\partial_t\alpha_n &=&-2e^{-2h|n|}\lp\beta_n-e^{\sign(n)h}\beta_{n-1}\rp \nonumber\\
\partial_t\beta_n &=&2e^{-2h|n|}\lp\alpha_n-e^{\sign(n)h}\alpha_{n+1}\rp.
\label{eq:eqs_lattice}
\end{eqnarray}
Now, we define the fields

\begin{equation}
\alpha_m=\sqrt{a}\alpha(x),\quad \beta_m=\sqrt{a}\beta(x), 
\end{equation}
where $a$ is the lattice spacing between the Dirac fermions $c_m$,
$x=ma$, which satisfy the usual anticommutation relations,
$\{\alpha(x),\alpha(x')\}=\{\beta(x),\beta(x')\}=2\delta(x-x')$ and
$\{\alpha(x),\beta(x)\}=0$. We find the continuum limit of the lattice
equations of motion by plugging these fields into
Eqs.\eqref{eq:eqs_lattice} and requiring $a\to 0$ and
$L\to\infty$ with both $\mathcal{L}=aL$ and $\hat{h}=h/a$ kept
constant,

\begin{eqnarray}
\partial_t\alpha(t,x) &\approx&-2ae^{-2\hat{h}|x|}\lp\partial_x-\sign(x)\hat{h}\rp\beta(t,x) \nonumber\\
\partial_t\beta(t,x) &\approx&-2ae^{-2\hat{h}|x|}\lp\partial_x-\sign(x)\hat{h}\rp\alpha(t,x),
\label{eq:eqs_continuum}
\end{eqnarray}
where we made the approximation  $e^{\sign(x)h}\approx(1+\sign(x)h)$. If
$h=0$ the equations of motion \eqref{eq:eqs_continuum} correspond to
the massless Dirac equation
$i\slashed{\partial}\Psi=(i\gamma^0\partial_0+i\gamma^1\partial_1)\Psi=0$,
where we have introduced the spinor

\begin{equation}
  \Psi^T=(\alpha(x^0,x^1),\beta(x^0,x^1)),\quad (x^0,x^1)=(2t,x).
  \label{eq:spinor}
\end{equation}
Our choice for the $\gamma$ matrices is $\gamma^0=-\sigma_2$,
$\gamma^1=i\sigma_3$ and $\gamma^3=\sigma_1$, where $\sigma_i$, $i=1$,
2, 3, are the Pauli matrices. Hereafter, we choose $a=1$ that
sets the Fermi velocity $v_F=1$, so we can simplify $\hat{h}=h$,
$\mathcal{L}=L$, and rewrite the equations of motion of the
inhomogeneous system, Eq. \eqref{eq:eqs_continuum}, as

\begin{equation}
  \lp-\sigma_2\partial_0+e^{-2h|x^1|}i\sigma_3\lp \partial_1-
  \text{sign}(x^1)h\rp\rp\Psi=0 \ . 
\label{eq:spinorial}
\end{equation}
The previous equation corresponds to the massless Dirac equation in a
curved spacetime whose metric depends on the inhomogeneity $h$, see
Appendix \ref{sec:curved} for details. The Dirac equation on a
generic metric can be written as Eq. \eqref{eq:dirac_curved},

\begin{equation}
\lp-\sigma_2\partial_0+\frac{i}{2}\omega_0^{01}\sigma_3+\frac{E_1^1}{E_0^0}\lp i\sigma_3\partial_1-\frac{i}{2}\omega_1^{01}\sigma_2\rp\rp\Psi=0,
\label{eq:dirac_spinorial}
\end{equation}
where $\omega_\mu^{ab}$ is the spin connection and $E_a^\mu$ is the
inverse of the {\em zweibein}. Comparing
Eq. \eqref{eq:dirac_spinorial} with our equations of motion
Eq. \eqref{eq:spinorial}, we obtain

\begin{eqnarray}
\frac{E^1_1}{E^0_0}&=& e^{-2h|x^1|},\\
\omega^{01}_0&=&-2e^{-2h|x^1|}h\text{sign}(x^1),\\
\omega^{01}_1&=&0.
\end{eqnarray}
The solution of these equations gives rise to the space-time metric:

\begin{eqnarray}
g_{00}=-e^{-4h|x|},\quad g_{11}=1 \  ,
\label{eq:metric_componets} 
\end{eqnarray}
whose Euclidean version is

\begin{equation}
ds^2=e^{-4h|x|}dt^2+dx^2=\Omega^2(x)dzdz,
\label{eq:metric}
\end{equation}
where $\Omega(x)=e^{-2h|x|}$ is the Weyl factor and

\begin{equation}
z=\tilde{x}+it, \quad \text{with}\quad
\tilde{x}=\int_0^x\frac{dy}{\Omega(y)}=\frac{\sign(x)}{2h}\(e^{2h|x|}-1\).
\label{eq:xtilde}
\end{equation}
The non-zero Christoffel symbols are

\begin{equation}
  \Gamma^0_{01}=-2h\sign(x), \quad \Gamma^1_{00}=-2h\sign(x)e^{-4h|x|}, 
\end{equation}
and the non-zero components of the Ricci tensor are

\begin{equation}
  R_{00}=-e^{-4h|x|}\lp4h\delta(x)-4h^2\rp, \quad R_{11}=4h\delta(x)-4h^2.
\end{equation}
The scalar of curvature $R=g^{\mu\nu}R_{\mu\nu}$ is

\begin{equation}
  R=8\lp h\delta(x)-h^2\rp,
\end{equation}
where we have used Eq. \eqref{eq:metric_componets}. Thus, $R$ is
singular at the origin and constant and negative everywhere else, thus
allowing for the holographic interpretation of the rainbow state that
has been discussed in the literature \cite{MacCormack.18}.

\subsubsection{Entanglement entropies} 
\label{ssub:entanglement_entropies}

\begin{figure}[h!]
\includegraphics[width=8cm]{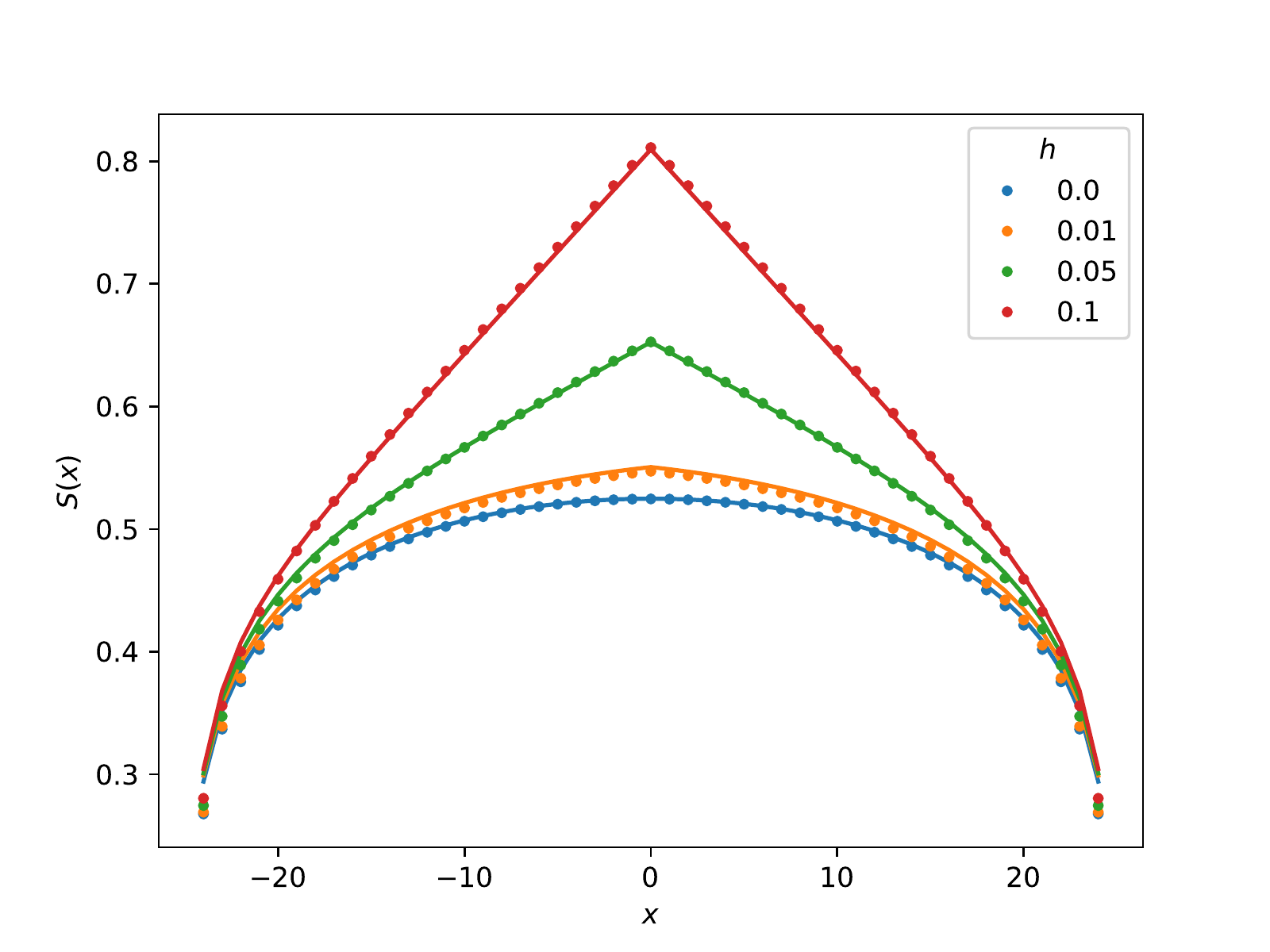}
\caption{EE of lateral blocks of the GS of Eq. \eqref{eq:ham_ising}
  with couplings \eqref{eq:hop_ising} for different values of $h$ and
  $L = 40$. The dots represent the numerical values obtained by exact
  diagonalization and the lines correspond to the predictions of
  Eq. \eqref{eq:ent_ising} using Eq. \eqref{eq:cprediction}.}
  \label{fig:1_ents_ising}
\end{figure}

We have shown that the continuum limit of the lattice model
Eq. \eqref{eq:major_ising} corresponds to a Majorana field in curved
space-time, described by a conformal field theory with central charge
$c=1/2$. We can obtain the EE of a block within this state employng
standard procedures \cite{Calabrese.04}, via the correlation
function of twist operators in an $n$-times replicated worldsheet
\cite{Laguna.17b}. The EE of the lateral blocks considered in the
previous section is

\begin{equation}
  S(x)=\frac{1}{12}\log\(\Omega(x)\frac{8\tilde{L}}{\pi}
  \cos(\frac{\pi \tilde{x}}{4\tilde{L}})\)+c'_I(\tilde{x}),
\label{eq:ent_ising}
\end{equation}
where the deformed quantities, $\tilde x$ and $\tilde L$, are computed
using Eq. \eqref{eq:xtilde}. The non universal function
$c'_I(\tilde{x})$ can be found using the relation between the EEs of
an XX chain of length $2L$ and an ITF chain of length $L$
\cite{Igloi.08}, and is given in Eq. \eqref{eq:cprediction} of
Appendix \ref{sec:igloi}. Fig. \ref{fig:1_ents_ising} shows the
numerical values of the EE for different values of $h$, showing the
agreement with Eq. \eqref{eq:ent_ising}. In the limit $h L \gg 1$,
Eq. \eqref{eq:ent_ising} implies for the half chain

\begin{equation}
 S(x=0)\approx \frac{1}{6}hL,
 \label{eq:thermal_ee} 
\end{equation}
which scales linearly with the system size, thus presenting a smooth
crossover between the weak and the strong inhomogeneity regimes for
which the EE is given by Eq. \eqref{eq:ent_rb}, i.e. $S_{A_L}=L\log2$.
In addition, this value of the EE can be interpreted as that of a
thermal state with an effective temperature $h/\pi$ \cite{Laguna.17b}.

\subsubsection{Entanglement Hamiltonian}

\begin{figure}
\includegraphics[width=8cm]{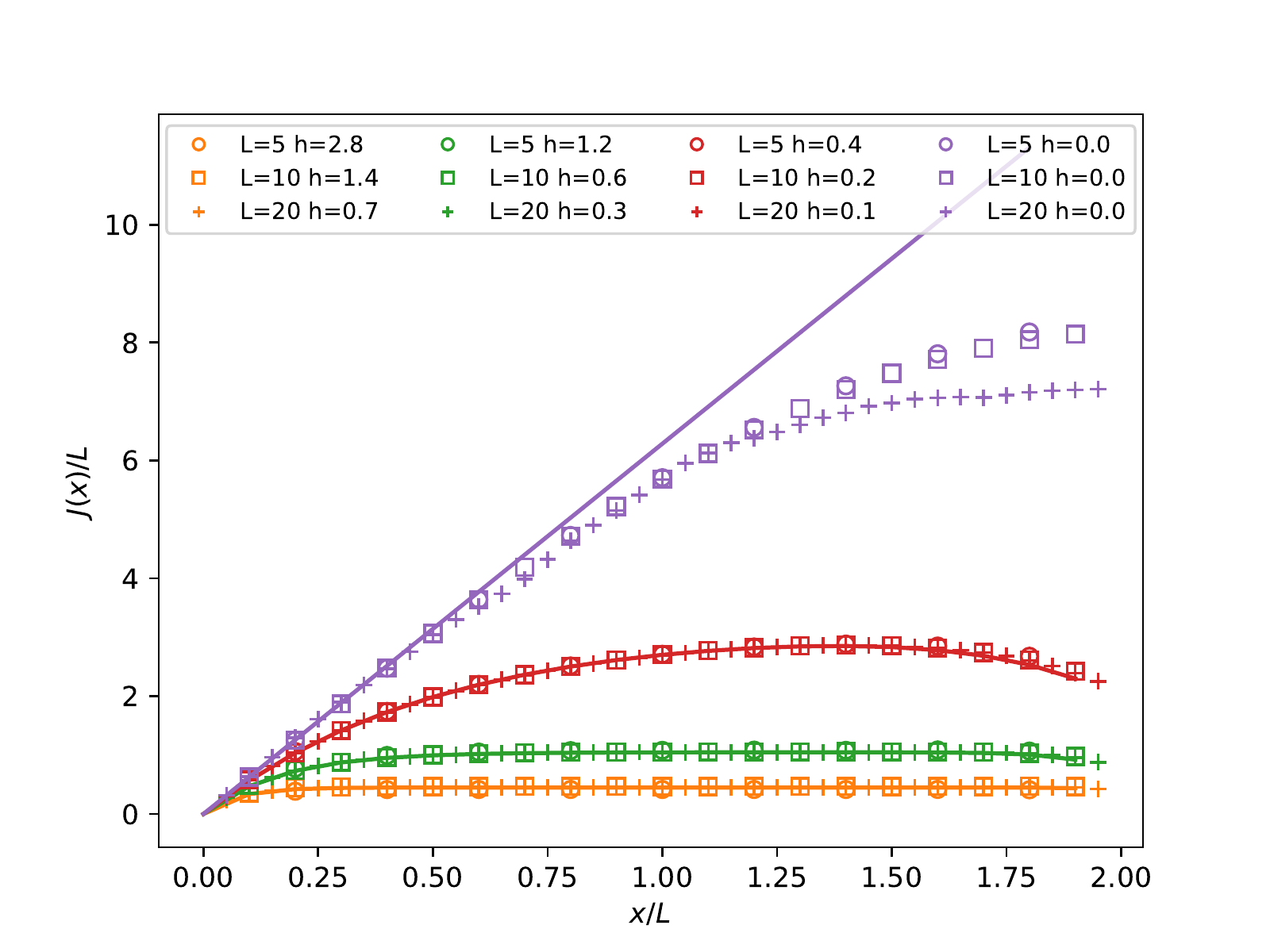}
\caption{Normalized weight functions $J(x)/L$ determining the EH for
  different values of $\lambda=2hL$. Continuous lines correspond to
  the theoretical prediction, Eq. \eqref{eq:profile_eh}. The purple
  straight line corresponds to the Bisognano-Wichmann prediction for a
  semi-infinite system.}
  \label{fig:eh_couplings}
\end{figure}

Let us now characterize the EH associated to the reduced density
matrix of the half chain, that we shall denote $\mathcal{H}_L$. In
Appendix \ref{sec:computing_eh} we discuss the standard procedure to
obtain the EH using the covariance matrix $C_L$
\cite{Peschel.03,Peschel.09}. The EH describes a local inhomogeneous
system with the weakest couplings near the center, which is the
internal boundary between the block and its environment. Moreover, if
the physical system is critical and infinite, it can be shown that
$\mathcal{H}_L$ is given by \cite{Cardy.16}

\begin{equation}
  \mathcal{H}_L=2\pi L\int_0^{L}dx\;J(x)\;T_{00}(x),
  \label{eq:eh_cft} 
\end{equation}
where $T_{00}(x)$ is the Hamiltonian density of the physical system
and $J(x)$ is a weight function. The Bisognano-Wichmann theorem
predicts $J(x)\approx x$ for a semi-infinite line \cite{BW.75,BW.76},
thus being approximately applicable for our case. Moreover, when the
original system is placed on a static metric, the weight function in
Eq. \eqref{eq:eh_cft} $J(x)$ should be appropriately deformed
following Eq. \eqref{eq:xtilde} \cite{Tonni.18}. In our case, we
obtain

\begin{equation}
  J(x)=\frac{2L}{\pi}\frac{e^\lambda-1}{\lambda}e^{-\lambda\frac{x}{L}}
  \sin\lp\frac{\pi}{2}\frac{e^{\lambda\frac{x}{L}}-1}{e^\lambda-1}\rp,
\label{eq:profile_eh}
\end{equation}
where $\lambda=2hL$. Near the internal boundary, which corresponds to
the center of the chain, the weight function $J(x)$ grows linearly
$J(x)\simeq 2\pi x$, as predicted by Bisognano and Wichmann
\cite{BW.75,BW.76}. Far from $\tilde{x}=0$, the weight function
develops a plateau, as it can be seen in Fig. \ref{fig:eh_couplings}
where $J(x)$ is plotted for different values of $\lambda$.

\section{Out of criticality}
\label{sec:noncrit}

Let us consider an inhomogeneous ITF model described by the Hamitonian
Eq. \eqref{eq:ham_ising} or, equivalently, Eq. \eqref{eq:major_ising},
with a modification of the coupling constants Eq. \eqref{eq:hop_ising}
studied in the previous section,

\begin{eqnarray}
J_m &=& \begin{cases}
e^{-2h|m+1/2|+\delta} & \text{if } m\neq -1/2, \\
e^{-h/2+\delta} & m=-1/2,\\ \end{cases}\\
\Gamma_m&=&e^{-2h|m|-\delta},
\label{eq:JGdeltas}
\end{eqnarray}
where $\delta\in\mathbb{R}$. Notice that if $h=0$ and $\delta\ll 1$,
then $J_m=1+\delta$ and $\Gamma_m=1-\delta$, and our system describes
a Majorana chain with alternating couplings, thus showing a relation
to the Kitaev chain and the Su-Schrieffer-Heeger (SSH) model
describing a dimerized chain of Dirac fermions
\cite{Su.79,Heeger.88}. Indeed, the parity term $e^{\pm\delta}$ pushes
the system described in Sec. \ref{sec:critical} out of criticality, as
we will describe throughout this section.

\subsection{Strong Inhomogeneity} 
\label{sub:strong_inhomogeneity_delta}

Let us consider the Hamiltonian given in
Eq. \eqref{eq:major_ising} in the limit $h\gg1$. We can apply
the same SDRG of the previous section, making use of the parameter

\begin{equation}
\kappa=\delta/h.
\label{eq:kappa}
\end{equation}
The RS is obtained when all the RG steps decimate the Majoranas at the
center of the chain, but we will show that other structures may be
obtained, depending on the value of $\kappa$. In order to decimate the
central pair we need $J_{1/2}$ to be the strongest coupling of the
chain. In other words, $J_{-1/2}>\Gamma_{1/2}$ which implies that
$e^{-h(1/2-\kappa)}>e^{-h(1+\kappa)}$. Hence, we arrive at the
condition

\begin{equation}
   \frac{1}{2}-\kappa<1+\kappa\Rightarrow \kappa>-\frac{1}{4}.
\end{equation}
Thus, if $\kappa>-1/4$ the Majorana fermions $\beta_{-1/2}$ and
$\alpha_{1/2}$ fuse into the Dirac fermion $b_{1/2}$, defined in
Eq. \eqref{eq:b12} and, using Eq. \eqref{eq:rg_G}, we obtain a
renormalized coupling

\begin{equation}
   \tilde{\Gamma}_{1/2}=e^{-3h\lp\frac{1}{2}+\kappa\rp},
\end{equation}
which will couple $\alpha_{-1/2}$ and $\beta_{1/2}$. These Majorana
fermions are decimated at the second RG step fusing into $d_{1/2}$,
Eq. \eqref{eq:d12}, if $\tilde{\Gamma}_{1/2}>J_{1/2}$, implying that 

\begin{equation}
   3\lp\frac{1}{2}+\kappa\rp<2-\kappa\Rightarrow \kappa<\frac{1}{8},
\end{equation}
and then a new term appears in the effective Hamiltonian of the form
$\tilde{J}_{3/2}\beta_{-3/2}\alpha_{3/2}$, where
$\tilde{J}_{3/2}$ follows from Eq. \eqref{eq:rg_J},

\begin{equation}
  \tilde{J}_{3/2}=e^{-5h\lp\frac{1}{2}-\kappa\rp}.
\end{equation}
Summarizing, the first central decimation requires $\kappa>-1/4$ while
the second requires $\kappa<1/8$. We can iterate this procedure and
find that the bound on $\kappa$ associated to exactly $n$ consecutive
central decimations is given by

\begin{align}
  \kappa &>-\frac{1}{4n},\text{ if $n$  odd,} \nonumber \\
  \kappa &<\frac{1}{4n}, \text{ if $n$ even.}
  \label{eq:kappa_bound}
\end{align}
The state with exactly $n$ central decimations will be called
$\ket{n}$. With this notation, the RS corresponds to $\ket{n=2L}$, and
satisfies 

\begin{equation}
\label{eq:kappars}
  |\kappa|<\frac{1}{8L}.
\end{equation}

\begin{figure*}
\includegraphics[width=150mm]{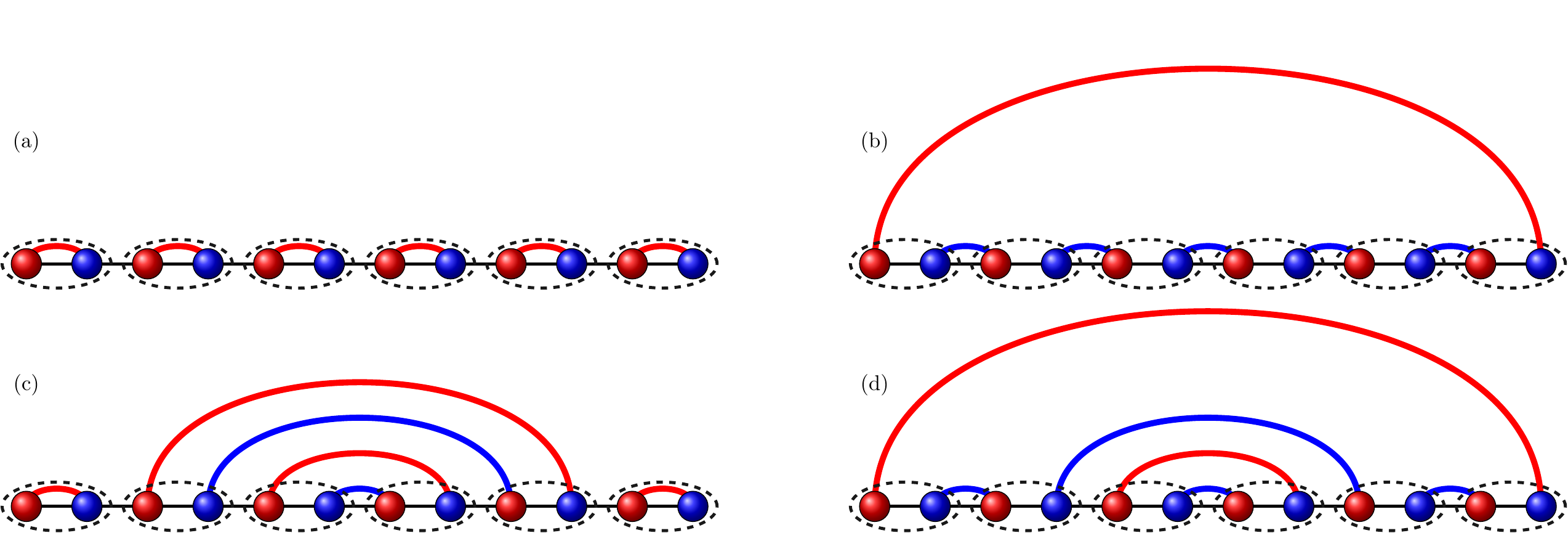}
\caption{Schematic representation of different ground states of
  Eq. \eqref{eq:major_ising} obtained with the SDRG scheme. The
  trivial (a) and non-trivial (b) local pairings correspond to the
  states $\ket{n=0}$ and $\ket{n=1}$, the two possible ground states
  of the Kitaev chain. Panels (c) and (d) illustrate ground states
  with a coexistence between non-local rainbow-like and local
  fermions. The topological type of the local pairing depends on the
  parity of $n$. Hence, we have a trivial pairing (c) $\ket{n=4}$ and
  a non trivial one $\ket{n=3}$ in (d).}
\label{fig:sdrgdimers}
\end{figure*}

Whenever a central decimation fails, the SDRG must choose the
strongest couplings between two identical links, symmetrically placed
with respect to the center of the chain. That is not a problem for the
algorithm, because the links are not consecutive \cite{Samos.20}. More
relevantly, from that moment on the RG will always proceed by {\em
  dimerizing} the chain towards the extremes, except perhaps for a
final long distance bond, depending on the parity of the system,
related to the Kitaev phase \cite{Kitaev.01}.

Thus, we are led to the following physical picture, which is
illustrated in Fig. \ref{fig:sdrgdimers}. In panel (a) we can see the
GS for $\kappa<-1/4$. No central bonds are created, and we obtain the
state $\ket{n=0}$. Panel (b) shows the GS for $\kappa>1/8$, in which a
single central bond is created. Due to parity reasons, a second bond
must appear between the extremes of the chain, thus leading to the
non-trivial Kitaev chain, which we call the state $\ket{n=1}$. Panel
(c) shows the state $\ket{n=4}$ and panel (d) the state $\ket{n=3}$,
which can be obtained within fixed ranges of $\kappa\in (-1/12,-1/20)$
and $\kappa\in(1/24,1/16)$ respectively, which can be found through
Eq. \eqref{eq:kappa_bound}.

\begin{figure*}
\includegraphics[width=80mm]{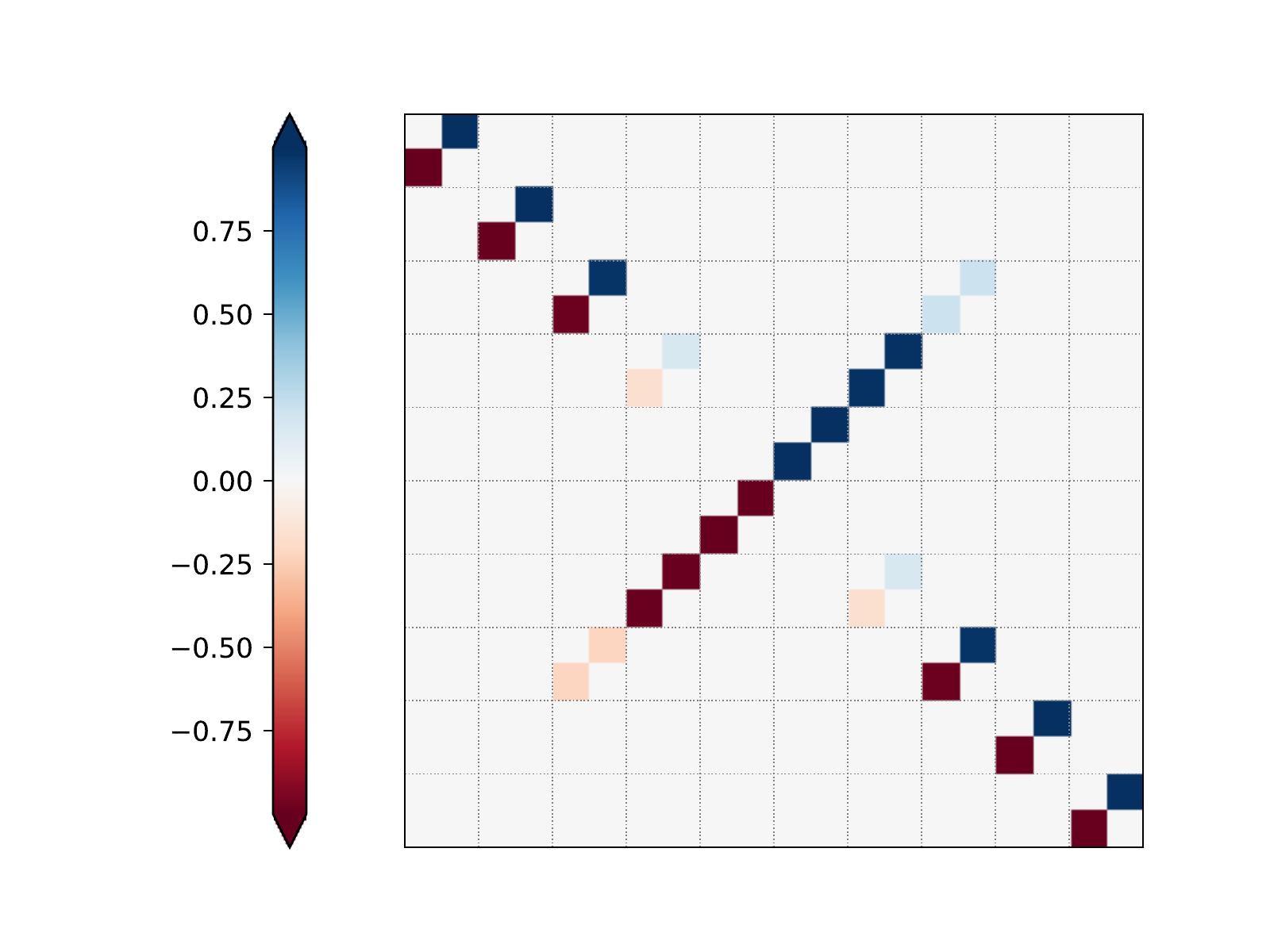}
\includegraphics[width=80mm]{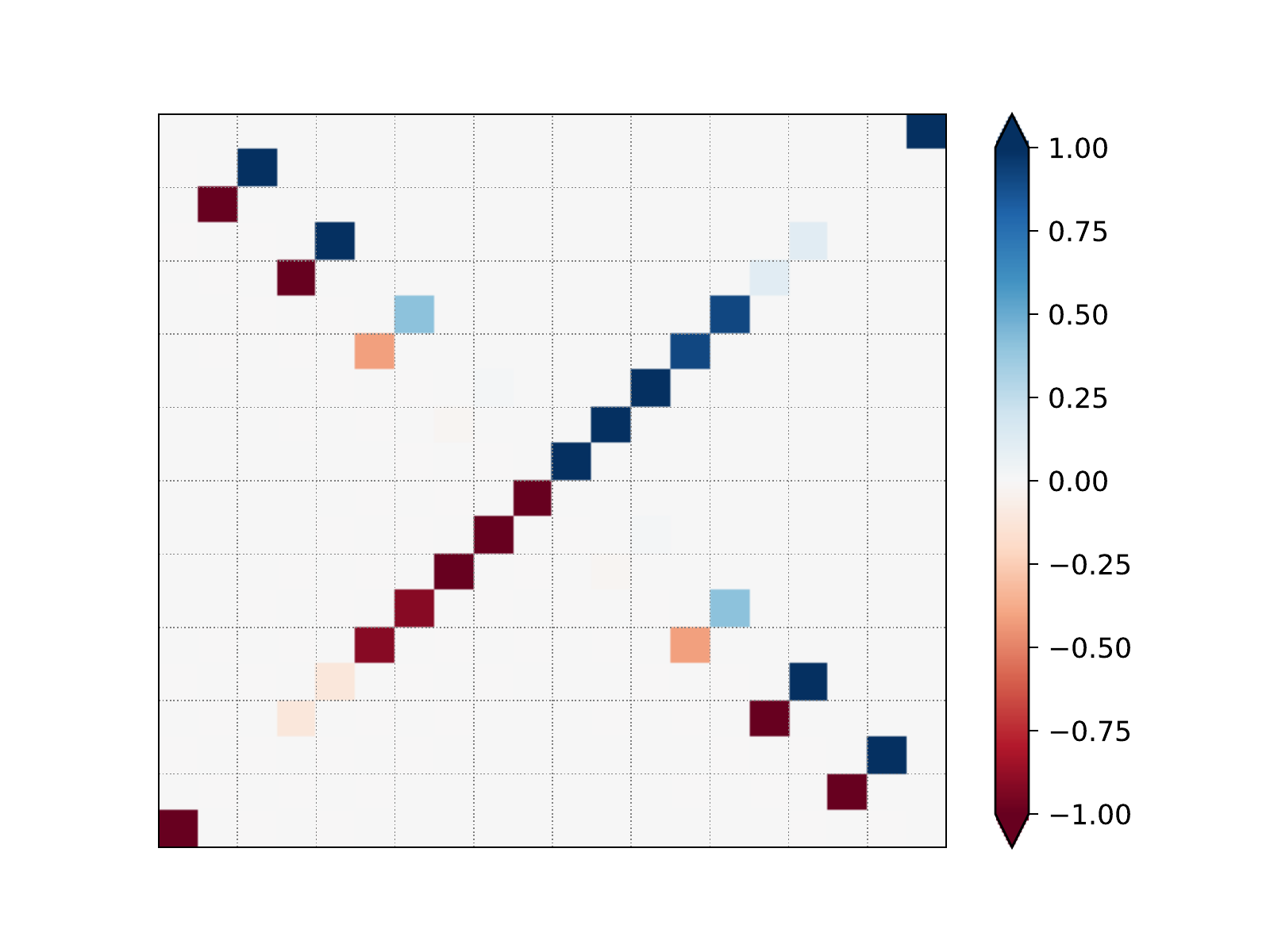}
\caption{Covariance matrices of states $\ket{n=4}$ and $\ket{n=5}$ for
  $h=10$ and $N=20$. The short-range single states populate the
  secondary diagonals while the long-range singlet states correspond
  to the antidiagonal.}
\label{fig:covariances}
\end{figure*}

This physical picture can be confirmed through the analysis of the
covariance matrices, which are depicted using a color code in
Fig. \ref{fig:covariances}. Indeed, we can see the CM for $N=20$ spins
and $h=10$, in the suitable range for $\ket{n=4}$ (left) and
$\ket{n=5}$ (right). The central patterns show $n=4$ and $n=5$ central
arcs, respectively. As predicted, the $n=5$ case presents an extra
bond between the extremes of the system, showing that it belongs to
the non-trivial Kitaev phase.

\begin{figure}
\includegraphics[width=80mm]{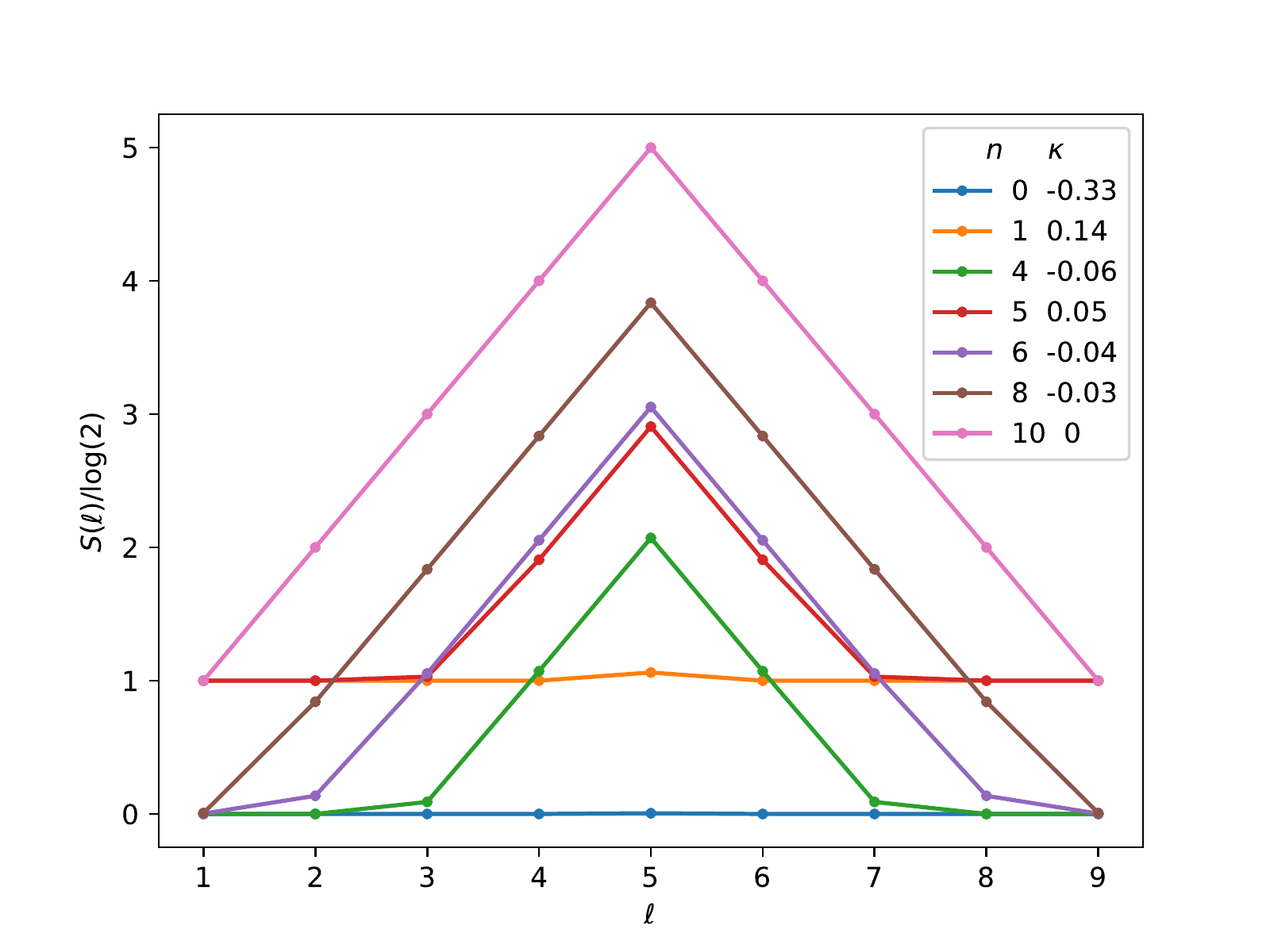}
\includegraphics[width=80mm]{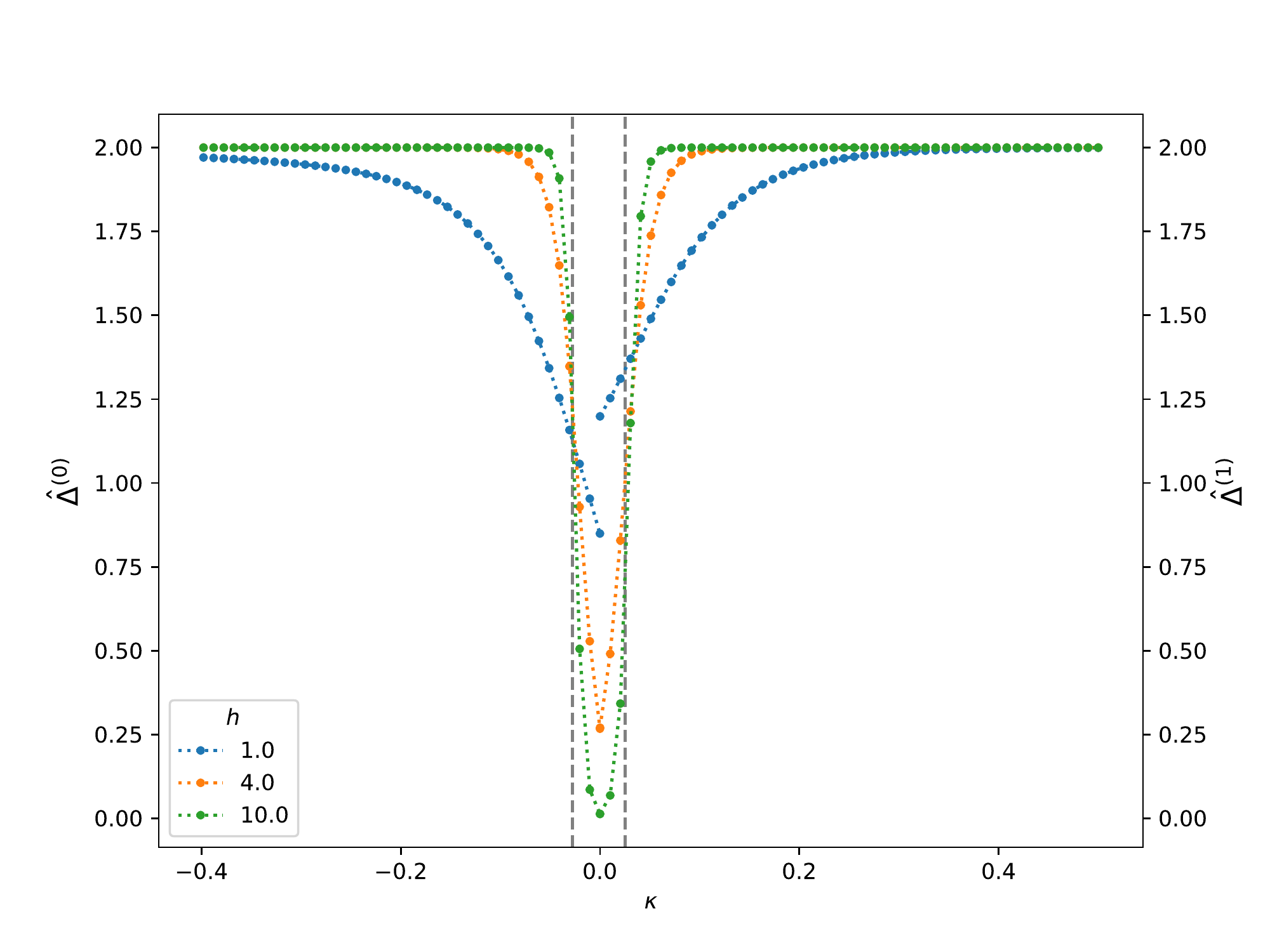}
\caption{Top: EE of lateral blocks of a chain of $N=10$ spins for
  different values of $\kappa$, using $h=10$. For $\kappa=-0.33$ and
  $\kappa=0.14$ we obtain respectively the $\ket{n=0}$ and $\ket{n=1}$
  states, for which the entropy is flat. On the other extreme, the
  $\kappa=0$ curve corresponds to the rainbow state, $\ket{n=10}$,
  which presents maximal EE growth. The intermediate values of
  $\kappa$ are chosen following Eq. \eqref{eq:kappa_bound} and present
  different numbers of central bonds. The EE for the half-chain,
  $\ell=5$, agrees with Eq. \eqref{eq:ent_recipe}. Bottom: scaled
  energy gaps ($\hat{\Delta}^{(0)}$ for $\kappa<0$ and
  $\hat{\Delta}^{(1)}$ for $\kappa>0$) of the same chain for different
  values of $h$. The vertical gray lines delimit the rainbow state
  region for this chain size ($n>2(L-1)$).}
\label{fig:ents_sdrg}
\end{figure}

\bigskip
 
Furthermore, we examine the EE of lateral blocks, $S(A_\ell)$ in the
top panel of Fig. \ref{fig:ents_sdrg}, which has been computed from
the CM using the same systems, with $N=10$ spins and $h=10$. As
predicted in our physical picture, the EE for the smallest block
begins at 0 or $\log(2)$ depending on the sign of $\kappa$, and
presents a linear tent-shape at the center, within a block of $\lfloor
n/2\rfloor$ spins and reaching an EE $n\log(2)/2$. The topological
nature of the states is clarified in Appendix
\ref{sec:pictorial_distinction_of_the_phases_}, where we present a
graphical way to distinguish the trivial and topological phases by
overlaying each state with the trivial state $\ket{n=0}$ and counting
the total number of loops.

\bigskip

We can also consider the energy gap around the Fermi level to capture
the differences between ground states. Defining the energy gap in an
inhomogeneous system presents some challenges, since it should be
expressed in units of the typical energy scale. A strategy that has
proved useful in similar cases is to rescale the energy gap with the
lowest coupling of the system \cite{Samos.20}, which in this case
becomes $\Gamma_{L-1/2}\approx e^{-2hL}$. Hence, the scaled gap

\begin{equation}
  \hat{\Delta}^{(0)} \equiv\frac{E_{N+1}-E_{N}}{\Gamma_{L-{1/2}}},
\end{equation}
becomes constant ($\hat{\Delta}^{(0)}=2$), as it can be seen in
Fig.\ref{fig:ents_sdrg} (b). The states $\ket{n \text{ odd}}$ present
a zero mode at the edge, and the system is strictly gapless,
$\hat{\Delta}^{(0)}=0$. Therefore, it is convenient to consider the
second gap, defined as

\begin{equation}
  \hat{\Delta}^{(1)} \equiv\frac{E_{N}-E_{N-1}}{J_{L-{1/2}}},
\end{equation}
which can also be seen in Fig. \ref{fig:ents_sdrg} (b). If
$n\leq(2L-3)$ there is a short-range Majorana singlet state and the gap
is finite, $\hat{\Delta}^{(1)}=2$. Yet, both gaps fall to zero for the
rainbow state, $\ket{n=N}$.

\subsection{Weak Inhomogeneity}
\label{sub:weak_inhomogeneity_delta}

Proceeding in the same way as in the previous section, we can obtain
the equations of motion from the Hamiltonian
Eq. \eqref{eq:major_ising} and describe the continuum limit
defining $x=am$, $a\to 0$, $h\to 0$, with $\hat{h}=h/a$ and
$\mathcal{L}=La$ kept constant, in terms of the fields $\alpha(x,t)$
and $\beta(x,t)$,

\begin{align}
\partial_t\alpha &\approx -2a e^{-2\hat{h}|x|} \Bigl(ae^\delta\partial_x-\Bigl(\sign(x)he^\delta +2\sinh\delta\Bigr)\Bigr)\beta,\nonumber\\
\partial_t\beta &\approx-2ae^{-2\hat{h}|x|} \Bigl( ae^\delta\partial_x-\Bigl( \sign(x)he^\delta -2\sinh\delta\Bigr)\Bigr)\alpha,
\label{eq:eqs_continuum_delta}
\end{align}
where we will use $a=1$ for convenience. These equations can be
rewritten in terms of a spinor field $\Psi$, Eq. \eqref{eq:spinor},
using the same $\gamma$ matrices, obtaining

\begin{align}
\Bigl(-\sigma_2\partial_0+e^{-2h|x^1|}\bigl(&i\sigma_3e^\delta\partial_1\nonumber\\
&-i\sigma_3\text{sign}(x^1)he^\delta +2i\sinh\delta\bigr)\Bigr)\Psi=0,
\end{align}
And, then, we can compare this equation with that representing the
dynamics of a Dirac field in a curved space time.

\begin{equation}
\(-\sigma_2\partial_0+\frac{i}{2}\omega_0^{01}\sigma_3+\frac{E_1^1}{E_0^0}\( i\sigma_3\partial_1-\frac{i}{2}\omega_1^{01}\sigma_2\)+i\frac{m}{E_0^0}\)\Psi=0.
\end{equation}
where $\omega^{ab}_\mu$ is again the spin connection and $E^\mu_a$ the
inverse of the {\em zweibein}. From the above identification we find that:

\begin{eqnarray}
  E^0_0&=&e^{2h|x^1|},\quad E^1_1=e^\delta\nonumber\\
  \omega^{01}_0&=&-2e^{-2h|x^1|}e^\delta h\text{sign}(x^1)\nonumber\\
  \omega^{01}_1&=&0,\nonumber\\
  m&=&2\sinh\delta,
  \label{eq:curved_parameters}
\end{eqnarray}
that leads to a (1+1)D metric whose non-zero terms are

\begin{equation}
  g_{00}=-e^{-4h|x|},\qquad g_{11}=e^{-2\delta}.
\end{equation}
However, $g_{11}\simeq 1$ if $\delta\ll1$, and thus the associated
metric coincides with the one found in the previous section, see
Eq. \eqref{eq:metric}. Thus, the field theory associated to the system
described by the Hamiltonian \eqref{eq:major_ising} is described by a
massive Majorana fermion, with $m\approx 2\delta$, placed in the
curved background described by the metric Eq. \eqref{eq:metric}.

\subsubsection{Entanglement entropies} 
\label{sub:ents_and_xi}

Let us first consider the case $h=0$, i.e. the massive fermion on a
flat space. The EE of this system has been obtained previously by
evaluating the associated two-dimensional classical model via the CTM
formalism \cite{Peschel.09}. For $\delta>0$ one obtains

\begin{equation}
  S(\delta)=\frac{1}{12}\(\log\(\frac{k^2}{16k'^2}\)+
  \(1-\frac{k^2}{2}\)\frac{4I(k)I(k')}{\pi}\)+\log2,
\end{equation}
while for $\delta<0$ we get

\begin{equation}
S(\delta)=\frac{1}{12}\(\log\(\frac{4}{kk'}\)
+\frac{1}{2}\(k^2-k'^2\)\frac{4I(k)I(k')}{\pi}\),
\label{eq:peschel_s}
\end{equation}
where $I(x)$ is the complete elliptic integral of the first kind
\cite{Abramowitz} and

\begin{equation}
  k=e^{-2|\delta|},\qquad k'=\sqrt{1-k^2}.
  \label{eq:k}
\end{equation}
Notice that if $|\delta|\ll1$, $k\approx(1-|\delta|)/(1+|\delta|)$
which is the value used in Refs. \cite{Peschel.09,Eisler.20}. Although
Eq. \eqref{eq:peschel_s} is only exact for the infinite chain, it is
still valid provided that $1/\delta\ll L$, i.e. when the cluster
decomposition principle is satisfied. Near the critical point,
$\delta\ll1$, Eqs. \eqref{eq:peschel_s} are simplified to

\begin{equation}
  S\approx\frac{c}{6}\log(\frac{1}{1-k})=\frac{c}{6}\log\xi,
\end{equation}
where the quantity inside the logarithm can be interpreted as a
correlation length $\xi$ \cite{Calabrese.04} with the appropriate
units of length,

\begin{equation}
  \xi=\frac{1}{1-k}\approx \frac{1}{2|\delta|},
  \label{eq:xi_C}
\end{equation}
which corresponds to the inverse of the mass, $m=2\delta$,
Eq. \eqref{eq:curved_parameters}. To end this brief summary of the
homogeneous non critical case, let us write the EE for the half chain
of a {\em finite} system as

\begin{equation}
  S(\delta,L)=\frac{c}{6}\log\frac{\xi_E(\delta,L)}{2}+b(\delta),
  \label{eq:ent_xiE}
\end{equation}
where $\xi_E(\delta,L)$ shall be called the {\em entanglement length},
because it plays the role of an effective correlation length in order
to compute the EE, even though its value is upper bounded by the size
of the system, $N=2L$. If $\delta=0$, the system is critical and
$\xi_E(0,L)$ saturates this bound, thus leading to the logarithmic
scaling predicted by CFT. On the other hand, if $|\delta|$ is large
enough then $\xi_E(\delta,L)\ll 2L$, finite size effects are not
important and the cluster decomposition principle holds. Thus, the
results for the infinite chain can be applied, and the area law is
satisfied. Hence, we see that in this case Eq. \eqref{eq:ent_xiE} is
just a reparametrization of Eq. \eqref{eq:peschel_s}.

Moreover, when we introduce inhomogeneity in the system through the
parameter $h$, we find that the EE can be obtained merely {\em
  deforming} the entanglement length $\xi_E(\delta,L)$ according to
the same prescription used before, given in Eq. \eqref{eq:xtilde},
giving rise to the Ansatz

\begin{equation}
  S(L,\delta,h)=\frac{c}{6}\log\(\tilde{\xi}_E(\delta,L)\)+b(\delta),
  \label{eq:ent_xi}
\end{equation}
where

\begin{equation}
  \tilde{\xi}_E(h,\delta,L)=\frac{1}{2h}\(e^{h\xi_E(\delta,L)}-1\),
  \label{eq:xitilde}
\end{equation}
is the deformed entanglement length, corresponding to the curved
space-time.

\begin{figure}[h!]
\includegraphics[width=80mm]{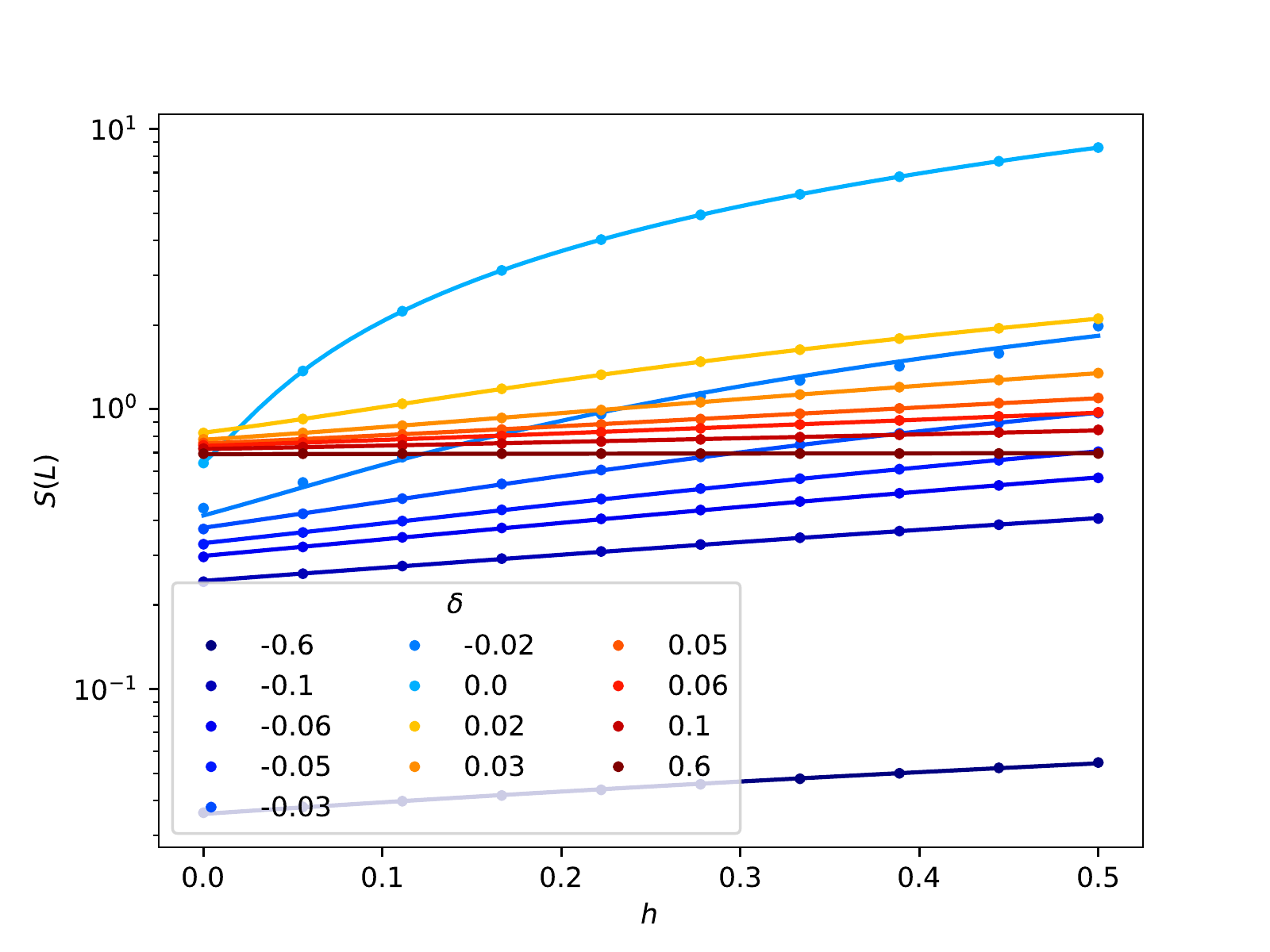}\\
\includegraphics[width=80mm]{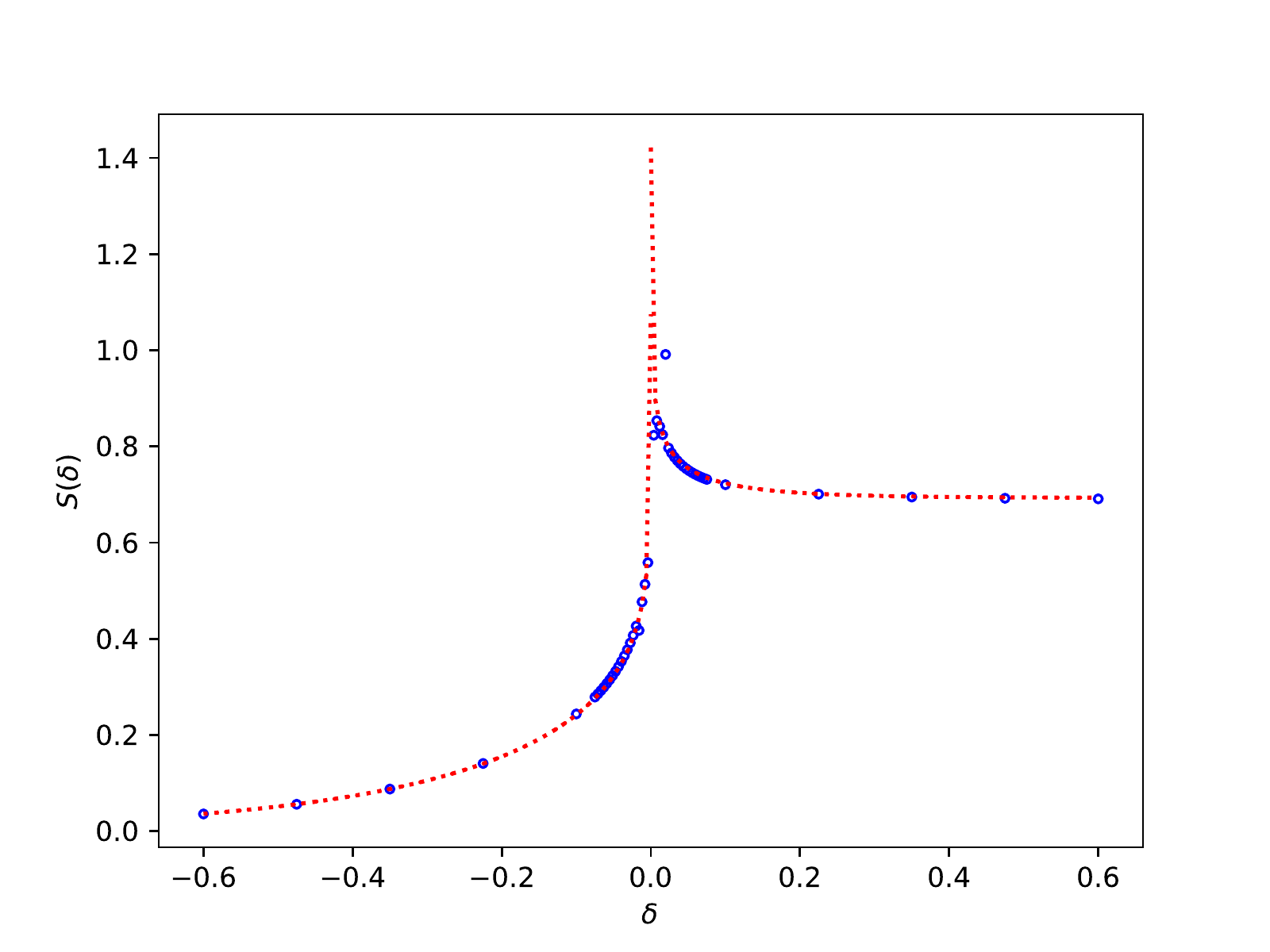}
\caption{Top: Plot of the fits (lines) performed with expression
  \eqref{eq:xitilde} of the numerical results for the half chain EE
  (points) for different values of $\delta$. Bottom: The dotted red
  line corresponds to Eq. \eqref{eq:peschel_s} and the blue dots to
  Eq. \eqref{eq:ent_xi}, where $\xi_E(\delta,L)$ has been obtained by
  fitting Eq. \eqref{eq:xitilde} to the numerical half chain EE of a
  system with $L=100$.}
\label{fig:fit_Peschel}		
\end{figure}

\begin{figure}[h!]
\includegraphics[width=80mm]{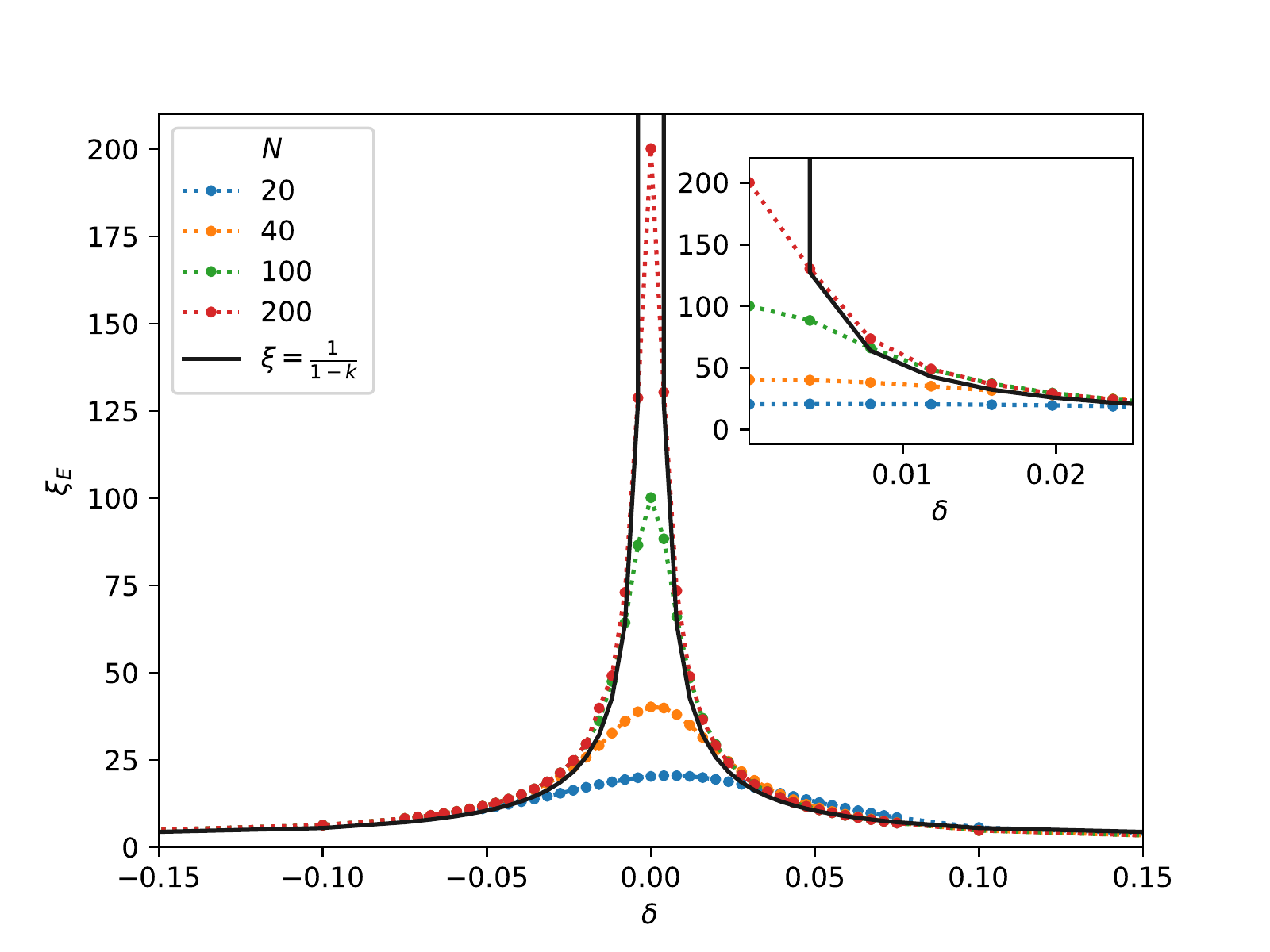}\\
\caption{Entanglement correlation length $\xi_E(\delta,L)$ for
  different chain sizes $2L$. Near the critical point we have
  $\xi_E\approx\xi$, Eq. \eqref{eq:xi_C}.}
\label{fig:corr} 
\end{figure}

We have fitted expression Eq. \eqref{eq:ent_xi} to the numerical
values for the EE of the half chain for different values of $\delta$
and $h$, using $\xi_E(\delta,L)$ and $b(\delta)$ as fitting
parameters. The agreement between the fits and the numerical results
can be seen in the top panel of Fig. \ref{fig:fit_Peschel}. Hence, we
obtain a single value for the entanglement length for each $L$ and
$\delta$, which accounts for the EE under different degrees of
inhomogeneity $h$. In the bottom panel of Fig. \ref{fig:corr} we can
see the good agreement between the infinite chain prediction,
Eq. \eqref{eq:peschel_s}, and the output of Eq. \eqref{eq:ent_xiE}
having used the values $\xi_E(\delta,L)$ and $b(\delta)$ that were
obtained from the previous fits.

In Fig. \ref{fig:corr} we present the fitted values $\xi_E(\delta,L)$
for different system sizes. The system presents universal behavior as
long as the correlation length is much smaller than the system size.

\bigskip

It is worth to ask whether the weak and strong inhomogeneity regimes
match smoothly. Let us consider the limit $h\gg1$ in
Eq. \eqref{eq:ent_xi},

\begin{equation}
  S(L,\delta,h)\approx\frac{c}{6}h\xi_E(\delta,L).
\end{equation}
If $\delta=0$ we have $S(L,0,h)\approx \frac{c}{3}hL$, as it was
discussed in the previous section. On the other hand, if $\delta\ll 1$
but $1/\delta\ll L$, $\xi_E(\delta,L)=\xi$ from
Eq. \eqref{eq:xi_C}. Hence,

\begin{equation}
  S(L,\delta,h)\approx\frac{h}{12\delta}=\frac{1}{12\kappa},
\end{equation}
which is a manifestation of the area law given by the interplay
between the inhomogeneity $h$ and the dimerization $\delta$. Thus, we
see that the weak and strong inhomogeneity regimes match.

\subsubsection{Entanglement Hamiltonian and Entanglement Spectrum}

\begin{figure}[h!]
\includegraphics[width=80mm]{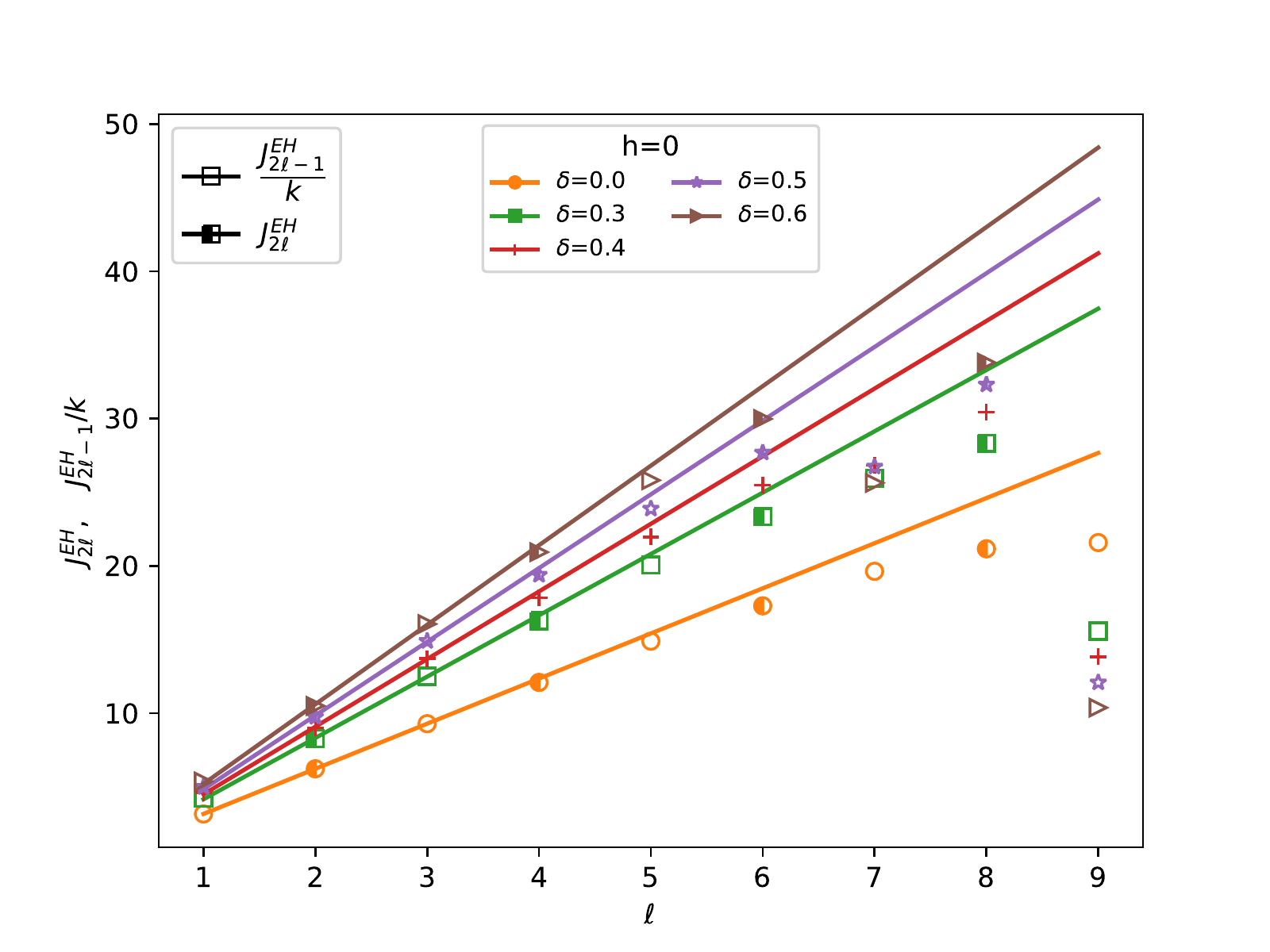}\\
\includegraphics[width=80mm]{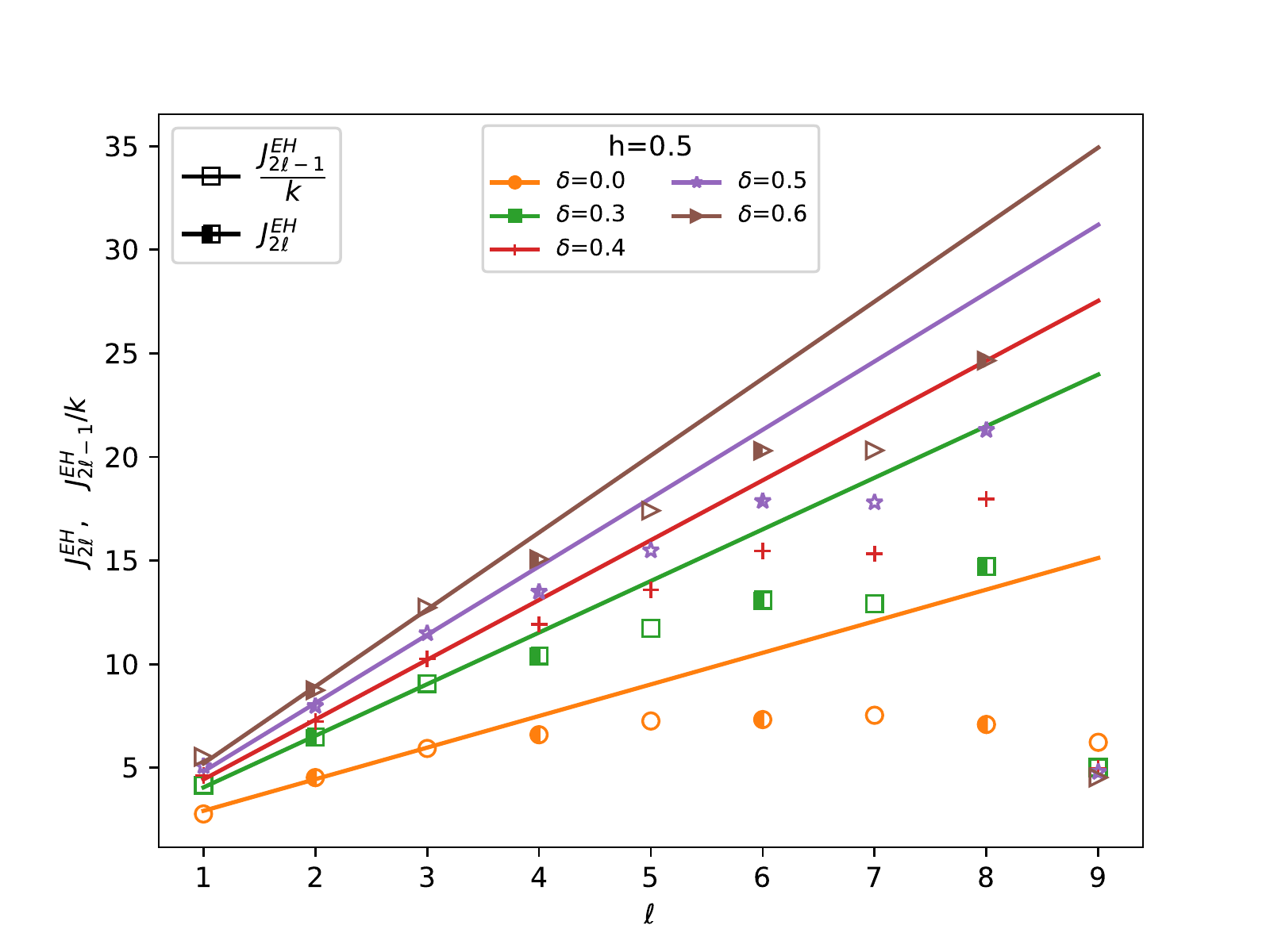}\\
\caption{Nearest neighbor couplings of a chain of $2N=20$ Majorana
  fermions, $J^{EH}_\ell$, with a parity compensation: odd values of
  $\ell$ are divided by $k$. (a) Case $h=0$. We can see that, despite
  the small size of the chain, we recover the behaviour of the
  infinite chain near the boundary for strong dimerizations. Empty
  symbols correspond to even couplings while the filled ones are for
  $J^{EH}_{2\ell-1}/k$, see Eq. \ref{eq:infinite_eh_ctm}. (b) In the
  case $h=0.5$ we observe the same initial behavior for low values
  of $\ell$.}
\label{fig:eh_hops} 
\end{figure}

The reduced density matrix $\rho_A$ of a half infinite chain can be
written in terms of the generator of the Baxter corner matrix,

\begin{equation}
  \rho_A=e^{-H_{CTM}},
\end{equation}
Since the model is integrable, we can simplify and state that
$H_{CTM}=\epsilon H_\mathcal{N}$, where $H_{\mathcal{N}}$ is a
Hermitian operator with integer spectrum. Thus, the ES $\epsilon_l$,
with $l=1\dots L$, is equally spaced and we may focus on the level
spacing $\epsilon$. For the ITF model we have

\begin{equation}
  \epsilon=\pi \frac{I(k')}{I(k)},
  \label{eq:level_spacing}
\end{equation}
where $k$ and $k'$ are given by Eq. \eqref{eq:k}. The EH of the half
infinite chain can be identified with the generator of the CTM
\cite{Eisler.20}. Thus, in the case of the ITF chain the first
neighbor couplings grow linearly from the internal boundary towards
the bulk with a parity oscillation between $1$ and $k$,

\begin{equation}
  \mathcal{H}=\sum_{l=1}^\infty
  J^{EH}_{2\ell-1}\alpha_\ell\beta_\ell+J^{EH}_{2\ell}\beta_\ell\alpha_{\ell+1},  
\end{equation}
with

\begin{align}
\displaystyle J^{EH}_{2\ell-1}=I(k')(2\ell-1),\quad
J^{EH}_{2\ell}=I(k')2\ell k,
\quad \delta<0\nonumber\\
\displaystyle J^{EH}_{2\ell-1}=I(k')(2\ell-1)k,\quad
J^{EH}_{2\ell}=I(k')2\ell,
\quad\delta>0 
\label{eq:infinite_eh_ctm}
\end{align}
where $\alpha_\ell$ and $\beta_\ell$ correspond to the lattice
Majorana fermions. Fig. \ref{fig:eh_hops} (a) shows the nearest
neighbor coupling constants of the EH, $J^{EH}_\ell$, slightly
modified in order to improve the visualization: for odd values of
$\ell$, $J^{EH}_\ell$ has been divided by $k$ in order to remove the
parity oscillation, leaving a linear growth with slope $2I(k')$, in
similarity to \cite{Eisler.20}. If we switch on the inhomogeneity,
setting $h=0.5$, we can observe the same EH couplings in
Fig. \ref{fig:eh_hops} (b): a linear increase of the couplings with a
parity oscillation beween values $1$ and $\hat{k}(h)$, which depends
on the inhomogeneity. Notice that $\hat{k}(0)=k$.

\section{Conclusions and further work}
\label{sec:conclusions}

In this work we have characterized the entanglement properties of an
inhomogeneous transverse field Ising critical spin-1/2 chain for which
both the couplings and external fields fall exponentially from the
center with a rate $h$, which defines the rainbow ITF model. It can be
analytically solved by mapping into a Majorana chain, which suggests
to treat the couplings and the external fields on an equal
footing. Applying the strong disorder renormalization method we find
that the ground state can be expressed in terms of generalized singlet
states which are displayed concentrically around the center, similarly
to the rainbow state. The weak inhomogeneity regime can be
characterized by taking the continuum limit and showing that the
resulting field theory corresponds to a free massless Majorana fermion
field on a curved space-time. Thus, we are able to predict the
behavior of the entanglement entropy deforming appropriately the known
CFT results for Minkowski space-time, which turns the characteristic
logarithmic growth into a linear growth with the block size. Moreover,
there is a smooth crossover between both regimes. The nearest neighbor
coefficients of the entanglement Hamiltonian present the standard
linear growth as we move away from the internal boundary, in agreement
with the Bisognano-Wichmann theorem, showing that the state can be
interpreted as a thermofield double for large enough inhomogeneity.

Out of criticality, we introduce a parity-dependent term whose
strength $\delta$ competes with $h$, attempting to destroy the linear
entanglement. Strong-disorder renormalization arguments show that for
each value of $\kappa\equiv \delta/h$ we obtain a fixed number of
concentric singlets around the center of the chain, also showing that
the trivial and non-trivial Kitaev phases are obtained for positive
and negative values of $\kappa$, although with a substantial
deformation. The weak inhomogeneity regime with small $\delta$ is
described by a massive Majorana field theory placed over the curved
space-time that we found in the case of the critical model. We have
computed the EE by defining an effective correlation length
$\xi_E(\delta,L)$ which is deformed with the metric, see
Eq. \eqref{eq:ent_xiE}. Near the entangling point, the entanglement
Hamiltonian presents a linear growth of the couplings with a parity
oscillation that can be accounted for using CTM results for the
infinite systems. The amplitude of the oscillation and the slope
depends on the inhomogeneity parameter.

In a previous work \cite{Samos.19}, we found a connection between the
rainbow antiferromagnetic Heisenberg spin chain with the Haldane
phase, and another between the rainbow XX spin chain and the BDI SPT
phases, by means of a folding transformation around the center of the
symmetry of the chain. It could be interesting to extend this approach
to the models considered in this manuscript and, more generally, to
address the entanglement characterization of inhomogeneous 2D
systems. In addition, it could be relevant to consider an experimental
realization of the rainbow state in terms of a Rydberg atoms chain
whose effective Hamiltonian is an inhomogeneous ITF model with an
additional longitudinal field \cite{Schauss.18}. It is possible to
extend Fisher's RG to this model and find the conditions under which a
rainbow is formed. Also, it could be interesting to consider strongly
inhomogeneous anyon models and study them harnessing their relation
with $SU(2)_k$ Chern-Simons theories \cite{Bonesteel.07,Nayak.08}.

\begin{acknowledgements}
We thank Gene Kim and Erik Tonni for conversations. We acknowledge the
Spanish government for financial support through grants
PGC2018-095862-B-C21 (NSSB and GS), PGC2018-094763-B-I00 (SNS),
PID2019-105182GB-I00 (JRL), QUITEMAD+ S2013/ICE-2801, SEV-2016-0597 of
the ``Centro de Excelencia Severo Ochoa'' Programme and the CSIC
Research Platform on Quantum Technologies PTI-001.
\end{acknowledgements}

\onecolumngrid

\appendix


\section{SDRG on Majorana chains}
\label{sec:dasgupta_majos}
In this appendix we explain the SDRG scheme applied to an inhomogeneous chain of Majorana fermions. Let us consider a system of $4$ Majorana fermions whose Hamiltonian is given by: 
\begin{equation}
\label{eq:4bodyham}
	H=i\lp g_{ia}\gamma_i\gamma_a+g_{ib}\gamma_i\gamma_b+g_{ab}\gamma_a\gamma_b+g_{aj}\gamma_a\gamma_j+g_{bj}\gamma_b\gamma_j\rp,
\end{equation}
with $\{\gamma_m,\gamma_n\}=2\delta_{mn}$. Let us assume that $g_{ab}$ is larger than the rest so we can use use perturbation theory to diagonalize \eqref{eq:4bodyham}.
\begin{align}
\label{eq:Hperts}
H_0&=ig_{ab}\gamma_a\gamma_b,\\
H_I&=i\lp g_{ia}\gamma_i\gamma_a+g_{ib}\gamma_i\gamma_b-g_{ja}\gamma_j\gamma_a-g_{jb}\gamma_j\gamma_b\rp \ . 
\end{align}
Defining the Dirac fermion $b=\frac{1}{2}\lp\gamma_a+i\gamma_b\rp$, we have that $H_0=2g_{ab}\lp b^\dagger b-1/2\rp$ whose spectrum is $\pm|g_{ab}|$ and eigenvectors $\ket{0},\ket{1}$ such that $b\ket{0}=0$. The $H_I$ can be written as:
\begin{equation}
\label{eq:Hpert}
	H_I=i\left\{\left[\lp g_{ia}-i g_{ib}\rp\gamma_i-\lp g_{aj}-i g_{bj}\rp\gamma_j\right]b+\left[\lp g_{ia}+i g_{ib}\rp\gamma_i-\lp g_{aj}+i g_{bj}\rp\gamma_j\right]b^\dagger\right\}
\end{equation}

Note that we must extend the Hilbert space:
$\ket{0}\to\ket{0}\otimes\ket{\psi}$ where $\ket{\psi}$ is an unknown
state of the Majorana fermions $\gamma_i,\gamma_j$. In the same way
$\ket{1}\to\ket{1}\otimes\ket{\varphi}$. However we shall make an
abuse of notation and write $\gamma_i$ instead of
$\mel{\psi}{\gamma_i}{\varphi}$. The first order corrections are zero
so we compute the second order:

\begin{equation}
  \Delta E_0 =\frac{\mel{0}{H_I}{1}\mel{1}{H_I}{0}}{E_0-E_1}.
\end{equation}
Using Eq. \eqref{eq:Hpert} we find the same corrections for $g_{ab}>0$
and $g_{ab}<0$

\begin{equation}
\Delta E_0= i\lp \frac{g_{ia}g_{bj}}{g_{ab}}-\frac{g_{ib}g_{aj}}{g_{ab}}\rp\gamma_i\gamma_j+\tilde{E},
\end{equation}
with $\tilde{E}_0=1/(2|g_{ab}|)\lp g_{ia}^2+g_{ib}^2+g_{bj}^2+g_{aj}^2\rp$. Thus, an effective Hamiltonian $H_{eff}=ig_{ij}\gamma_i\gamma_j$
emerge with the hopping term given by

\begin{equation}
  g_{ij}=\frac{g_{ia}g_{bj}}{g_{ab}}-\frac{g_{ib}g_{aj}}{g_{ab}}.
\end{equation}
Particularizing for the ITF spin chain, $g_{ib}=g_{aj}=0$ and we
recover Eq. \eqref{eq:rg_J} and \eqref{eq:rg_G}. In
Ref. \cite{Bonesteel.07} the authors provide the same result with a
graphical derivation. The matrix elements are computed by counting the
loops obtained by overlaying the (generalized) singlet states. See
Appendix \ref{sec:pictorial_distinction_of_the_phases_} for details of
the overlaying procedure.

\section{Covariance Matrices}
\label{sec:covariance}

Let us consider a system of $N=2L$ Majorana fermions
given by the quadratic Hamiltonian:

\begin{equation}
  \mathcal{H}=i\bm{\gamma}^TA\bm{\gamma},
  \label{eq:ham_generic}
\end{equation}
where $\bm{\gamma}^T=(\gamma_1\dots\gamma_N)$ and $A=-A^T$. There
exists a transformation $Q\in SO(2N)$ which brings the Hamiltonian to
the canonical form

\begin{equation}
  Q^TAQ=N_\epsilon,
\end{equation}
where $N_\epsilon$ is a block diagonal matrix

\begin{equation}
  N_\epsilon=\left(
\begin{array}{c|c} 
  0 & \bm{\epsilon} \\ 
  \hline 
  -\bm{\epsilon} & 0 
\end{array} 
\right),
\label{eq:neps1}
\end{equation}
where $\bm{\epsilon}$ is an $L\times L$ matrix with
$\text{diag}(\bm{\epsilon})=(\epsilon_1,\dots,\epsilon_L)$, which are
the positive eigenvalues of the matrix $iA$. We have then that

\begin{equation}
  \mathcal{H}=i\bm{\gamma}^TA\bm{\gamma}=i\bm{\gamma}^TQN_\epsilon Q^T\bm{\gamma}=i\tilde{\bm{\gamma}}^TN_\epsilon\tilde{\bm{\gamma}},
  \label{eq:ham_spinor_tilde}
\end{equation}
where we have defined a set of Majorana fermions
$\tilde{\bm{\gamma}}=Q^T\bm{\gamma}$. These Majorana fermions can be
arranged into Dirac fermions
$\bm{\Psi}^T=(\bm{b},\bm{b}^\dagger)=(b_1\dots b_L,b^\dagger_1\dots
b_L^\dagger)$ with $\bm{\Psi}=U\tilde{\bm{\gamma}}$ and

\begin{equation}
  U=\left(
\begin{array}{c|c} 
  \mathbb{I} & i\mathbb{I} \\ 
  \hline 
  \mathbb{I} & -i\mathbb{I} 
\end{array} 
\right),
\end{equation}
where $\mathbb{I}$ is the $L\times L$ identity matrix. The Hamiltonian
takes the form $\mathcal{H}=\sum_k\epsilon_k(b^\dagger_k b_k-1)$ and
the correlators are particularly simple,

\begin{equation}
  \expectationvalue{\bm{\Psi}\bm{\Psi}^\dagger}=\left(
\begin{array}{c|c} 
  \mathbb{I} & 0 \\ 
  \hline 
  0 & 0 
\end{array} 
\right).
\end{equation}
We can express it back in terms of the Majorana fermions
$\expectationvalue{\bm{\Psi}\bm{\Psi}^\dagger}=U\expectationvalue{\tilde{\bm{\gamma}}\tilde{\bm{\gamma}}^T}U^\dagger$
and those in terms of the physical Majorana fermions $\bm{\gamma}$,

\begin{equation}
  \expectationvalue{\bm{\gamma}\bm{\gamma}^T}= Q^T\left(
\begin{array}{c|c} 
  \mathbb{I} & i\mathbb{I} \\ 
  \hline 
  -i\mathbb{I} & \mathbb{I} 
\end{array} 
\right)Q.
\end{equation}
The symmetric part of the matrix above is given by the anticommutation
relation of the Majorana fermions, while the antisymmetric part that
contains all the non-trivial information is known as the covariance
matrix.

\section{Dirac Fermion in Curved Spacetime}
\label{sec:curved}

Let us consider the Dirac equation in curved spacetime:

\begin{equation}
	(i\slashed{D}-m)\Psi=0,
	\label{eq:dirac_eq}
\end{equation}
where $\slashed{D}=E^\mu_a\gamma^a D_\mu$ is the slashed covariant
derivative, and $E^\mu_a=g^{\mu\nu}\eta_{ab}e^b_\nu$ is the inverse
of the tetrad basis (or {\em zweibein}) $e^a_\mu$ that satisfies
$g_{\mu\nu}=e^a_\mu e^b_\nu\eta_{ab}$. More precisely, the covariant
derivative of the two component spinor $\Psi$ is given by

\begin{equation}
 D_\mu\Psi=\lp\partial_\mu-\frac{1}{8}\omega^{ab}_\mu[\gamma_a,\gamma_b]\rp\Psi,  	
\end{equation}
where $\omega^{ab}_\mu$ is the spin-connection which is defined in
terms of the Christoffel symbols $\Gamma^\nu_{\sigma\mu}$ and the
inverse of the tetrad $E^\mu_a$,

\begin{equation}
  \omega^{ab}_\mu=
  e^a_\nu\partial_\mu E^{b\nu}+e^a_\nu E^{b\sigma}\Gamma^\nu_{\sigma\mu},    	
\end{equation}
As we are considering a static system, it is reasonable to assume that
the tetrad matrix $e^a_\mu$ is diagonal ($E_0^1=E_1^0=0$). Expanding
\eqref{eq:dirac_eq} with this assumption leads to

\begin{equation}
	\lp iE_0^0\gamma^0\lp\partial_0-\frac{1}{8}\omega_0^{ab}[\gamma_a,\gamma_b]\rp+iE_1^1\gamma^1\lp\partial_1-\frac{1}{8}\omega_0^{ab}[\gamma_a,\gamma_b]\rp-m\rp\Psi=0 \ .
\end{equation}
Taking in account that
$[\gamma_0,\gamma_1]=[-\gamma^0,\gamma^1]=-2\gamma^3$ and the
antisymmetry of the internal indices of the spin connection we arrive
at

\begin{equation}
\lp\gamma^0\partial_0+\frac{1}{2}\omega_0^{01}\gamma^0\gamma^3+\frac{E_1^1}{E_0^0}\lp\gamma^1\partial_1+\frac{1}{2}\omega_1^{01}\gamma^1\gamma^3\rp+i\frac{m}{E_0^0}\rp\Psi=0 \ . 
\label{eq:dirac_curved}
\end{equation}

\section{Non universal function of the EE}
\label{sec:igloi}

The relation between the entanglement entropies of the XX and ITF
models is given by \cite{Igloi.08}:

\begin{equation}
S_{\text{XX}}(2x,2L)=2S_{\text{ITF}}(x,L).
\label{eq:Igloi}
\end{equation}
We will compute the non universal part of the EE
\eqref{eq:ent_ising} with the above expression. The EE of the XX model
whose ground state is a RS has been studied in the past
\cite{Laguna.17b},

\begin{eqnarray}
S_{XX}(2x,2L) &=& S_{cft}(2x,2L)+\frac{\gamma_1}{2}+S_{oscl}(2x,2L), \\
S_{cft}(2x,2L)&=&\frac{1}{6}\log\(e^{-h|2x|}\frac{8(e^{h2L}-1)}{h\pi}\cos(\frac{\pi(e^{h|2x|}-1)}{2(e^{h2L}-1)})\)\nonumber\\
S_{oscl}(2x,2L)&=&(-1)^{2x+2L}\(\frac{8(e^{h2L}-1)}{h\pi}\cos\(\frac{\pi(e^{h|2x|}-1)}{2(e^{h2L}-1)}\)\)^{-1},\nonumber
\end{eqnarray}
and $\gamma_1\approx0.4950+1/3\log2$ \cite{Jin.04}. Hence, by using
relation Eq. \eqref{eq:Igloi} we have

\begin{eqnarray}
c'_I(x,L)&=&\frac{\gamma_1}{4}+\frac{1}{6}\log2+\nonumber \\
&(-1)^{L}&\(\frac{16(e^{2hL}-1)}{h\pi}\cos\(\frac{\pi(e^{2h|x|}-1)}{4(e^{2hL}-1)}\)\)^{-1}.
\label{eq:cprediction}
\end{eqnarray}

\section{Computation of the Entanglement Hamiltonian}
\label{sec:computing_eh}
The entanglement Hamiltonian can be obtained by knowing the covariance matrix. Consider a
system of $N=2L$ Majorana fermions given
by the quadratic Hamiltonian Eq. \eqref{eq:ham_generic}. 
There exists a transformation $O\in SO(2N)$ which brings the
Hamiltonian to the canonical form $O^TAO=N'_\epsilon$, where
$N'_\epsilon$ is a block diagonal matrix

\begin{equation}
N'_\epsilon=\bigoplus_k^L\lp\begin{matrix}0 & \epsilon_k\\ -\epsilon_k & 0\end{matrix}\rp,
\label{eq:neps2}
\end{equation}
where $\pm\epsilon_k, k=1\dots L$ are the eigenvalues of the matrix
$iA$. Notice that $N'_\epsilon$, Eq. \eqref{eq:neps2}, and
$N_\epsilon$, Eq. \eqref{eq:neps1}, are similar matrices meaning that
$O$ and $Q$ differ in the order of the elements of the basis. The
transformation $O$ is more convenient because the lateral blocks
considered in the main text are contiguous in this basis. Thus, the
Hamiltonian reads also as Eq. \eqref{eq:ham_spinor_tilde} after
substituting $Q$ by $O$. The density matrix $\rho$ associated to the
the GS of a quadratic Hamiltonian can always be written as

\begin{equation}
  \rho=\mathcal{K}e^\mathcal{-H},
\end{equation}
where $\mathcal{K}$ is a normalization constant and $\mathcal{H}$ is
called the entanglement Hamiltonian (EH), given by
\eqref{eq:ham_generic}. It is possible to obtain $\mathcal{H}_A$, the
EH associated to the reduced density matrix $\rho_A$ by knowing the
associated partial covariance matrix
$\mathcal{C}_A=\expectationvalue{\gamma_i\gamma_j}$ with $i,j\in A$.

\begin{equation}
  \mathcal{C}_A=O^TN'_\lambda O,
\end{equation}
$N'_\lambda$ has the same structure as $N'_\epsilon$ but contains the
eigenvalues of $\mathcal{C}_A$. Since the matrix $O$ brings to the
normal form both $\mathcal{C}_A$ and $\mathcal{H}_A$, there is a
relation between the eigenvalues of the EH $\epsilon$, known as
entanglement spectrum (ES), and the eigenvalues of the covariance
matrix:

\begin{equation}
  \lambda_k=-\tanh\frac{\epsilon_k}{2}
\end{equation}
Hence, by inverting the above relation, it is possible to compute the
EH knowing the covariance matrix,

\begin{equation}
  \mathcal{H}=O^TN'_{\epsilon(\lambda)}O.
  \label{eq:eh_recipe}
 \end{equation}

\section{Pictorial distinction of the topological phases} 
\label{sec:pictorial_distinction_of_the_phases_}

The trivial and topological ground states $\ket{n}$ of
Eq. \eqref{eq:major_ising} can be distinguished graphically. We start by overlapping the GS with the
trivial Majorana singlet state $\braket{n}{n=0}$ and connecting the
Majorana fermions (red and blue balls) with their opposites, leading
to the formation of closed loops.

\begin{figure}[t]
\centering
\includegraphics[width=150mm]{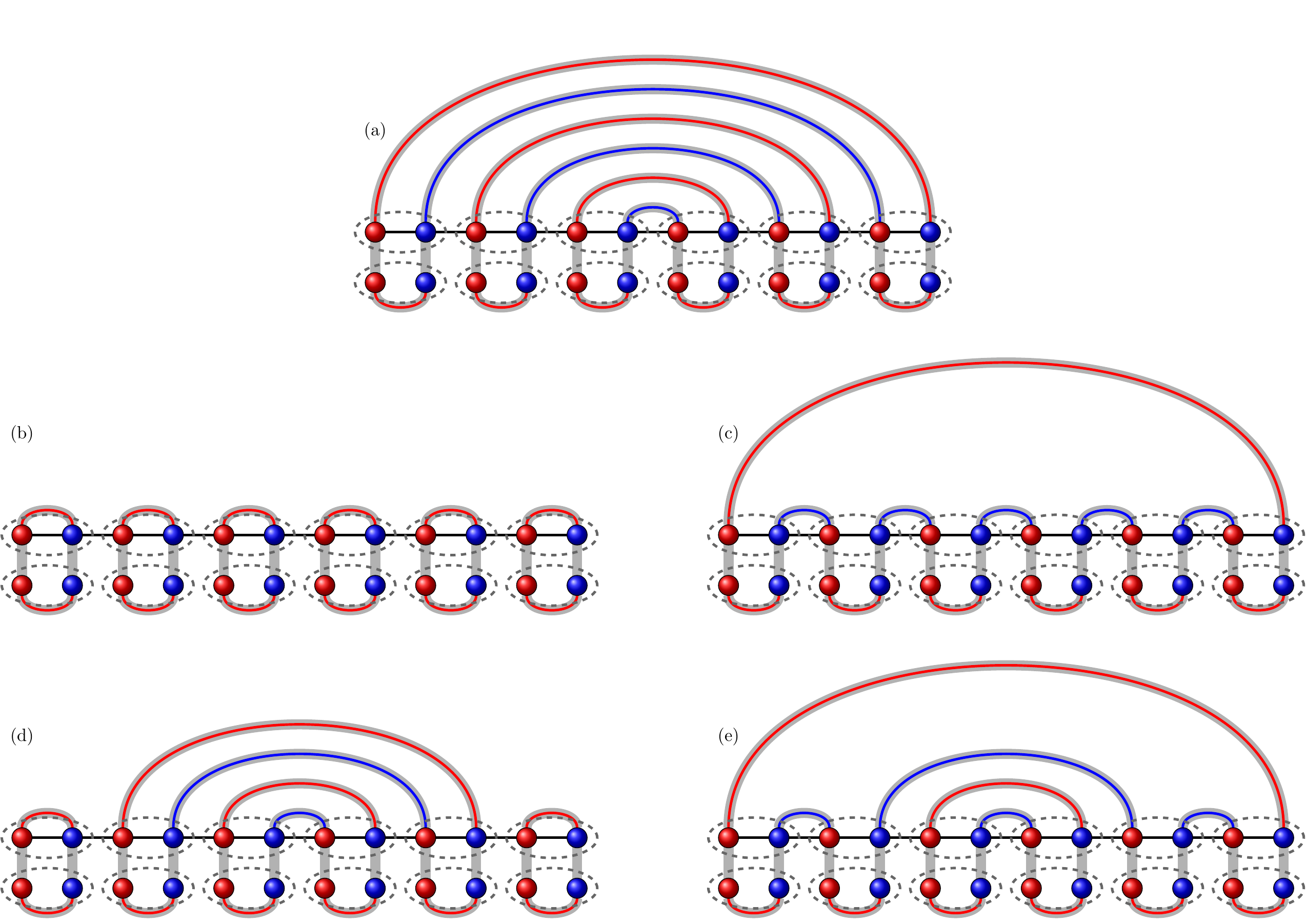}\\
\caption{Overlapping of the same states considered in
  Fig. \ref{fig:sdrgdimers}. Discussion on the main text.} 
\label{fig:overlaying}
\end{figure}

In Fig. \ref{fig:overlaying} we show the same GS that we presented in
Fig. \ref{fig:sdrgdimers} overlapping with $\ket{n=0}$, which we will
call $\braket{n=0}{n=0}$, that leads to $N$ loops matching with the
$N$ Dirac fermions of kind $c$, see Eq. \eqref{eq:cfermions}. This can
be seen in panel $(b)$. On the other side, the overlapping
$\braket{n=1}{n=0}$ leads to just one big loop as it can be seen in
panel $(c)$. Thus, the topological phase is characterized by a big
loop that encloses all the Majorana fermions. Considering $1<n<L-2$
central decimations, the overlapping $\braket{n=2m}{n=0}$ with
$m=0,\cdots,L-1$ decreases the total number of loops to $N-m$ while
the overlapping $\braket{n=2m-1}{n=0}$ with $m=1,\cdots,L-1$ increases
them up to $m$. For instance, in Fig. \ref{fig:overlaying} $(d)$ we
see the overlapping $\braket{n=4}{n=0}$ that leads to $6-2=4$
bonds. In panel $(e)$ there are $2$ loops, because the overlapping
corresponds to $\braket{n=3}{n=0}$. Finally, as it can be seen in
panel $(a)$, the overlaying of a rainbow state $\ket{\text{RS}}{n=0}$
leads to $N/2$ loops. This is another way of unveiling the criticality
of the RS since it corresponds to the intermediate situation.

The loops can also be interpreted in terms of spins and Fisher's RG
\cite{Bonesteel.07}. Each loop contains those spins that were
hybridized in consecutive RG steps with dominant $J$. Hence, the state
$\ket{n=1}$ is a {\em superspin} while the RS can be seen as a
collection of hybridized pairs of spins. However, notice that they do
not form a singlet state as it is the case of the RS of the XX chain
Eq. \eqref{eq:rsxx}.

\end{document}